\documentclass[twocolumn,showpacs]{revtex4}

\usepackage{amsmath, amssymb}
\usepackage{dcolumn}
\usepackage{latexsym}

\usepackage{indentfirst}
\usepackage{amssymb}
\usepackage{tabularx}
\usepackage{graphicx}

\begin{document}

\title{Evidence of two effects in the size segregation process in a dry granular media}

\author{G. F\'elix $^1$ and N. Thomas $^2$}

\altaffiliation[Corresponding author: ]{IUSTI, CNRS-UMR 6595, 5 rue E. Fermi, Technop\^ole de Ch\^ateau-Gombert, 13453 Marseille, France. \\ Tel: (+33) 4 91 10 68 72; Fax: (+33) 4 91 10 69 69.\\} 
\email{nathalie.thomas@polytech.univ-mrs.fr}

\affiliation{
$^{1}$ Laboratoire Magmas et Volcans, CNRS-UMR 6524,\\
5 rue Kessler, 63 000 Clermont-Ferrand, France\\
$^{2}$ IUSTI, CNRS-UMR 6595, 
5 rue E. Fermi, \\ Technop\^ole de Ch\^ateau-Gombert, 13453 Marseille, France}

\begin{abstract}
In a half-filled rotating drum, the {\it size segregation of particles of equal density} builds a ring pattern of the large particles, whose location continuously varies from the periphery to the center depending on the size ratio between particles [Thomas, Phys. Rev. E 62, 1 (2000) 961-974].
For small size ratios (typically$<$5): large beads are at the surface of the flow, as usually observed; for high size ratios (typically$>$15): large beads are close to the bottom in a reversing position. The existence of circles with an intermediate radius shows that the segregation at an intermediate level  within a flow is possible. 
In this paper, we experimentally study the segregation of particles of {\it different densities and sizes} in a half-filled rotating drum and other devices (chute flow, pile). In the drum, the location of the segregated ring continuously varies from the periphery to the center and is very sensitive to both the size (from 1 to 33) and density (from 0.36 to 4.8) ratios. The densest large beads segregate on a circle close to the center, the lightest large beads on a circle close to the periphery. Consequently, we found that for any tracer, its excess of mass, due to only a size excess, a density excess, or both, leads to a deep inside segregation of the tracer. There is a push-away process that makes heavy beads of any type go downwards, while the excess of size is already known to push large beads towards the surface, by a dynamical sieving process. Each segregation at an intermediate ring corresponds to a balance between these mass and geometrical effects. The segregation level in the flow is determined by the ratio of the intensities of both effects. 

\end{abstract}
\maketitle

\section{Introduction}
When a granular matter composed of several types of particles is put into motion, a de-mixing of the different particles occurs. This segregation appears whenever there are differences in size, density, shape, roughness...  For example, in a mixture of large and small beads of the same matter, the large beads are often found at the surface of the bed after the granular matter has been put into motion \cite{savagelun} or vibrated \cite{rosato}, or at the periphery of a half-filled rotating drum \cite{cantelaube95,clement95,dury97}. Segregation of beads of different sizes has been interpreted as a dynamical sieving during the flow which makes the small particles go down and, consequently, the large ones end up at the surface. This process comes from the size difference and is actually a geometrical effect. However, this interpretation is incomplete because it does not take into account the fact that large beads may push away their neighbors to go downwards \cite{tanaka, moi}, and can get buried into the bed.

\begin{figure}[t!]    
\center
\includegraphics[width=7.0cm]{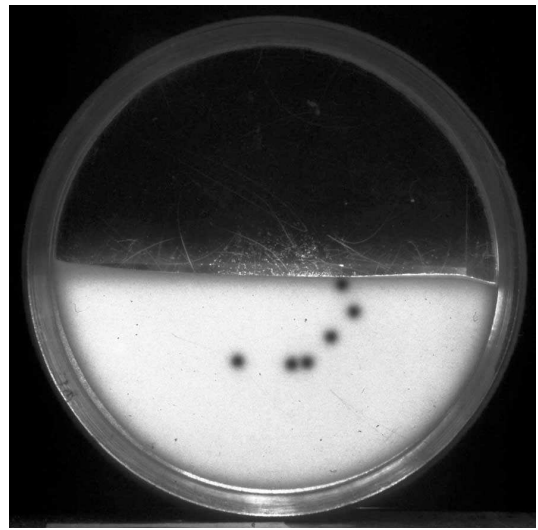}
\caption{Ring of segregation of 1.5-mm large tracer beads (Glass) in a bed of 90-$\mu$m small Glass beads, in a quasi 2D (2 dimensions) drum. The radial location of tracers is precisely defined (note that the upper bead is still in the flowing layer). The position is very sensitive to the value of the size ratio \cite{moi} and of the density ratio [this paper]. The variation of the ring radius is a powerful means to study the geometrical effect of segregation by making an equivalence with the variation of position induced by a change of density.}
\label{anneau}
\end{figure}

In a previous paper \cite{moi}, we investigated the balance between the two effects implied in the size segregation, named the geometrical and mass effects: the geometrical sieving pushes the large beads to the free surface, while the mass effect pushes the large (and heavy) beads to the bottom. This balance produces surprising rings of segregation [Fig. \ref{anneau}].
In three experimental devices (flow in an inclined channel, formation of a heap, half-filled rotating drum) a mixture of large and small Glass beads has been sheared during the flow. For a small fraction of large beads (typically 3\%), the segregated positions of the large beads continuously evolve from the surface to the bottom when increasing the size ratio, giving the up, intermediate and reverse segregations. Typically, for size ratios between 1 and 5, large beads are visible at the surface, and for size ratios above 15, they are close to the bottom. This evolution is masked when using a large fraction of large beads (for example 50\%). We concluded that, for small size ratios, the geometrical sieving is dominant while for larger ratios, the mass effect is dominant. But, for intermediate size ratios, the large beads are found at an intermediate level inside the bed: the 2 effects exactly counterbalance at this level, whose location depends on the value of the size ratio. 
In a rotating drum, this intermediate segregation produces rings of segregated tracers [Fig.~\ref{anneau}] 
whose location can be accurately measured, and is very sensitive to the size ratio between the beads. The position of the ring is then closely related to the relative intensities of the geometrical and mass segregation effects. In most previous studies, it is only possible to know if segregated particles are at the surface or at the bottom of the flow, {\it i.e.} there are only two possible patterns of segregation (or three when counting also the homogeneous state). But contrary to these classical patterns, the evolution of the position of the ring is continuous: there is an infinite number of patterns. Consequently, it is possible to continuously quantify the relative intensity of the segregation effects by measuring the ring radius. 

The aims of the present study are (1) to confirm that, in the size segregation of beads of equal density, a mass effect is responsible of the reverse and intermediate positions of large beads \cite{moi} and (2) to obtain data to quantitatively separate the two effects, geometrical and mass, both implied in the size segregation of beads of equal density. 
For that reason, we vary the mass effect independently of the geometrical effect, by changing the tracers density. We then investigate the segregation of beads of different densities, for each size ratio. 

Experimentally, the segregation due to a density difference has been less studied than the size segregation, one reason being the difficulty in varying the density. Some studies focus on phenomena happening during the shaking of a mixture \cite{breu, nagel} or follow the trajectories of tracers during the vibration \cite{duran, shinbrot}. The previous experimental studies involving a granular matter sheared into a slow frictional motion, report that for equivalent sizes, the dense particles are found at the bottom of the 
bed (or center of the drum), and the light particles at the surface (or periphery of the drum) \cite{cantelaubethese, khakhar97, khakhar03}. But they concern large density ratios (from 3 to 7). Numerical studies \cite{ristow94, khakhar97, McCarthy, khakhar99chute, khakhar03} use a continuous range of density. Some results are close to a ring segregation and show a continuous evolution of the segregated tracer location versus the mass ratio \cite{ristow94}. When using particles of different sizes and densities, the two differences act in addition or in competition, leading to up, down or no segregation \cite{dury99, metcalfe}. For example, a difference of size (ratio 3) can dominate a small difference of density (ratio 1.5): large dense particles are on surface \cite{Hill}. 
Experimentally \cite{alonso, metcalfe} or numerically \cite{dury99}, adjusting densities and sizes can lead to a homogeneous bed, interpreted as a global balance between density and size effects. But these studies concern only the segregation state (homogeneous or not) of the whole granular matter because they use similar fractions of large and small particles, contrary to the segregation pattern in rings which gives both the segregation intensity (ranging from homogeneous to totally segregated) {\it and the location} of the segregated minority particles. 

The present study concerns two sets of experiments where particles are submitted to a frictional shear motion during the flow (no vibration, no collisional flow).  

In the first part, we precisely measured the ring segregation pattern in a half-filled rotating drum. 
The segregation process is mainly considered as a one-particle process as shown by \cite{cantelaube95, clement95}, and results of all identical tracers regrouping at the same preferential level. A few large tracer beads of different densities (density ratios from 0.36 to 4.8) have been placed in a bed of small Glass beads whose size varies (size ratios from 1 to 33). 
Small tracers volume fractions (0.16\% to 3\%) are used; consequently, there is no percolation process, even for large size ratios. Large beads are considered as tracers, but are numerous enough to allow statistics (10 to 50 beads). We are aware of the fact that interactions between large beads modify the process at larger fractions \cite{moi} even if most features are still qualitatively valid, but we do not want to treat that point here. The aim is to relate the ring position to the forces that are exerted on each tracer for the tracer to reach a particular equilibrium level in the flow. In that way, we can study the forces responsible for the segregation, knowing (1) the location where they balance, and (2) how this varies with size and density ratios.

In the second part, three experimental devices have been used to show that the reverse and intermediate segregations are not only characteristic of the drum device: formation of a pile, half-filled rotating quasi 2D disc, flow along a slope. The general behavior of the segregation (up-inside-homogeneous) has been determined as a function of both density (ratios from 0.5 to 3.2) and size ratios (ratios from 0.2 to 10). Finally, the link between the two parts is made through values of tracer rings.

\section{Experiments in a rotating drum}\label{drum}
The drum is half-filled and turns around its axis, placed horizontally. The slow frictional flowing layer is located on the free surface. The remainder of the beads rotates in a solid way with the drum. An initially homogeneous mixture of two types of beads is obtained by placing the tracers by hand between successive thin layers of Glass beads, the drum axis being vertical during the filling. Two de-mixing processes take place in a drum: radial \cite{clement95, cantelaube95, moi} and axial segregations \cite{dasgupta91, zik94, Khosropour00}. The axial segregation is the longer one to appear, and we avoid it by stopping the rotation after a few revolutions. 

We follow the behavior of the minority component (tracers, indexed 2) in a bed of the majority component (small Glass beads, indexed 1). A mixture is characterized by diameter $d_2/d_1$ and density $\rho_2/\rho_1$ ratios. This study is limited to the case of large tracers ($d_2/d_1$$\geqslant$1). 

\begin{figure}[htbp]
\center
\includegraphics[width=5.5cm]{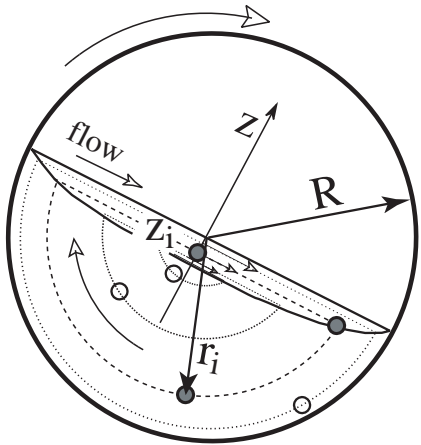}
\caption{Each radial position $r_i$ in the rotating solid part corresponds to a `vertical' (perpendicular to the free surface) location $z_i$ in the flowing layer.}
\label{zetr}
\end{figure}

We interpret the radial segregation as resulting from a segregation process happening within the flowing layer. The three patterns in the drum (tracers at the periphery, on a ring, at the center) are respectively due to the three locations of the segregation in the flow (tracers at the surface, at an intermediate level, at the bottom). 
The segregation is quantified by measuring the distances $r_i$ from the center of tracers to the center of the drum [Fig.~\ref{zetr}]. 
Two parameters are used: \newline(1) the spreading of the $r_i$ distribution, related to the segregation intensity and measured by the standard deviation of the $r_i$ distribution. When it is low, the intensity is strong, and the segregation pattern is well focused. When it is large, the segregation is weak and tracers are widely spread over the bed.\newline
(2) $r$ the average of $r_i$, defined as the ring radius. $r/R$ indicates the mean location of the segregated tracers whether the segregation is strong or weak ($R$ is the drum radius). Each radius $r_i$ corresponds to a `vertical' level $z_i$ in the flowing layer [Fig.~\ref{zetr}] \cite{chakraborty}. Roughly $r_i$=0 comes from tracers whose center is at the bottom of the flow, and $r_i$=$R$ from tracers whose center is at the free surface (half-embedded). The link could be done by the shape of the interface between static and flowing zones, which can be approximated by a parabolic function when the free surface is plane \cite{Khakhar97b, gwenthese}. But, we observed that beads whose centers are above the free surface present larger $r_i$ than beads just half-embedded, although a simple intersection linking $z_i$ and $r_i$ would give a saturation to the maximal value for all these beads. We assume that a tracer may stop when its bottom touches the interface between flowing layer and static zones: $r_i$ would then be smaller than the radius corresponding to the intersection between $z_i$ and the interface. This remark is based on qualitative observations, and we do not have data for a $z_i$-$r_i$ link. Moreover, tracers have a finite size and $r_i/R$ maximum is equal to 1$-$$d_2/2R$ (0.938 for 3-mm tracers), when the tracer touches the periphery. Due to the flow thickness, $r_i/R$ minimum is not equal to 0 either. As the densest tracers are never visible during the flow, we assume that the flow thickness is at least $d_2$. We do not take into account the beads located at less than 1.5$d_2$ from the free surface, because they could be touching the flowing layer when the rotation stops. $r_i/R$ is then superior to $1.5 d_2/R$ (0.18 for 3-mm tracers).

\subsection{Experimental device}
The drum is 42~mm long and 48.5~mm in diameter, made of Glass with Steel caps, and half-filled with a mixture of small Glass beads and large tracer beads (Table~\ref{billes}). The variation of the density ratio is obtained by changing the tracer density, the variation of the size ratio by changing the size of the small Glass beads. Tracers, usually 3~mm in diameter, represent from 0.16\% to 3\% of the beads volume and do not interact enough for their position to be affected \cite{moi}. Lead and Glass tracers are nearly spherical with the dimensions of each tracer comprised in the given range (Table~\ref{billes}) and there is no noticeable shape effect on the segregation. Small Glass beads are very-well sorted, having less than 10\% of relative size spreading \cite{moi}, and whose diameters range from 90~microns to 4~mm: 90; 150; 180; 200; 212; 250; 300; 355; 400; 425; 500; 600; 710; 850; 1000; 1500; 2000; 2500; 3000; 4000~$\mu$m. 
The Glass density is 2.5~g/cm$^{3}$, measured with a picnometer from 2.46 to 2.54 
with no systematic variation with the size. Some experiments using small Zirblast beads (500~$\mu$m, 3.85~g/cm$^{3}$) for the bed have also been done. 
\begin{table}
\caption{Tracers used in the drum experiments.}
\begin{center}
\begin{tabular}{|l|c|c|}
\hline
  ~~~~ \underline{Materials :}& \underline{density \small{(g/cm$^{3}$)}}&\underline{sizes \small{(mm)}}\\
Polypropylene \small{(PP)}&0.9&3\\
Polyamide \small{(PA})&1.14&3\\
Polyacetal \small{(POM)}&1.41&3\\
Teflon \small{(PTFE)} &2.15&3\\
Glass$^*$&2.5&3-3.14  \\
Aluminum &2.7&2.5 and 3.97\\
Silicone Nitride \small{(Si$_3$N$_4$)}&3.16&3 and 3.175\\
Ceramic  \small{(Al$_2$O$_3$)}&3.9&3\\
Titane &4.52&3\\
Zircone  \small{(ZrO$_2$)}&6.0&2\\
\small{ZrO$_2$} \footnotesize{(only Zirblast exp.)}&6.0&3\\
Steel &7.85&1, 2.5 and 3\\
Steel \footnotesize{(only Zirblast exp.)}&7.91&3\\
Lead$^*$&12&3.02-3.07 \\
\footnotesize{$*$not perfectly spherical} &    &  \\ 
\hline
\end{tabular}
\label{billes}
\end{center}
\end{table}

The humidity in the laboratory room is held at 50-55\% and special care is given to experiments involving the smallest beads to minimize electrostatic and humidity cohesive effects \cite{fraysse}. But electrostatic effects are strong when using Teflon beads which attract small Glass beads. 

Rotation speeds correspond to a continuous flow and to a plane free surface: around 0.085~s$^{-1}$ (from 0.04 to 0.1~s$^{-1}$) and up to 0.15~s$^{-1}$ in few cases [Figs.~\ref{r/Racetal}, \ref{r/Rverre}, \ref{r/RZrO2}, \ref{r/Racier}]. Experiments with a bed made of the largest Glass beads (resp. 2-2.5-3-4~mm) have been systematically done with higher speeds to obtain a continuous flow (resp. 0.1-0.12-0.13-0.2~s$^{-1}$). There is no difference between rotation obtained by a motor and by manually rolling the drum, which is the case for a third of the experiments, with no systematic order compared to size or density ratios.

After 3 to 6 revolutions, we carefully dip the drum into water. We can open it without disturbing the spatial distribution of the beads. For the largest bed beads, the drum is frozen before openning it. Several cross-sections are carefully made perpendicular to the drum axis, to measure the radial position $r_i$  of each tracer [Fig.~\ref{zetr}]. We noticed that tracers touching the caps have a slightly different mean location from tracers inside the bed, and removed them for the measurements.

\subsection{Circles of segregation}
\begin{figure}[htpb]  
\center
\includegraphics[width=7.8cm]{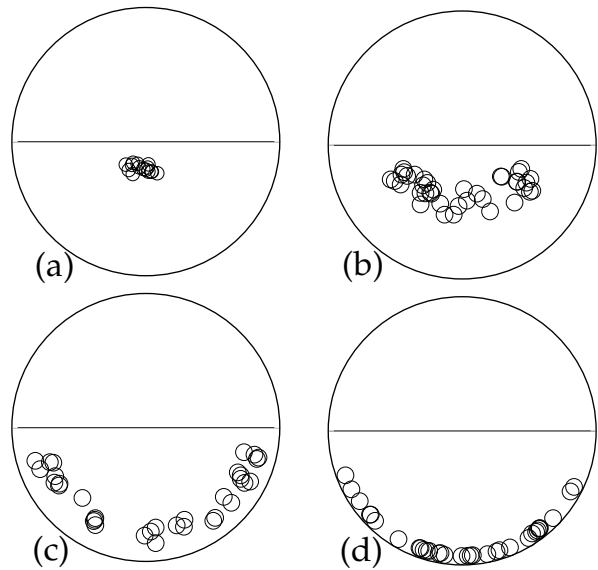}
\caption{Tracer locations in a bed of small Glass beads. Several cross-sections from one single experiment have been added on each drawing. Tracer material \& Glass beads diameter are: (a) Steel (2.5~mm) \& 300~$\mu$m, (b) Al$_2$O$_3$ (3~mm) \& 710~$\mu$m, (c) Teflon (3~mm) \& 400~$\mu$m, (d) Polyacetal (3~mm) \& 600~$\mu$m.}
\label{cercles}
\end{figure}

After segregation, the large beads are found on a ring whose radius varies with the mixture characteristics [Fig.~\ref{cercles}]. After 1 revolution, the ring radius is constant when the drum continues to turn \cite{moi}, corresponding to the stable segregation pattern obtained after one passage in the flowing layer \cite{vanpuy}. We then rotate 3 turns for most of our experiments. In a 2D drum \cite{dury97}, the convergence towards the stable pattern is less rapid and also shows an increase of the convergence time to the final segregated state at high rotation speeds. As our measurements of convergence were done in experiments at `low speed', for the highest speeds the drum is rotated 6 turns, rather than 3, to take into account a possible increase of the convergence time.

Figures \ref{r/Rpropyl} to \ref{r/Rplomb} represent the radius of the segregation ring versus the beads diameter ratio, for growing tracer densities. Lines are provided to highlight data trends \cite{fit}. Each mean radius $r$ is associated to an `error bar' which is the standard deviation of the $r_i$ distribution.

\subsubsection{Intensity of the segregation}\label{artefact}
\begin{figure}[htpb]
\center
\includegraphics[width=7.4cm]{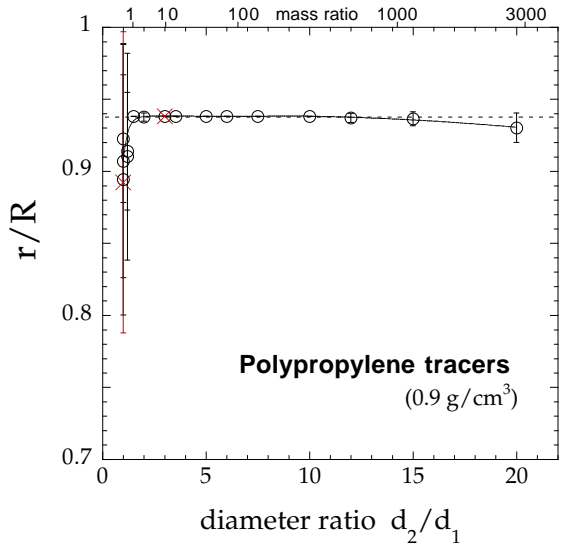}
\caption{Mean locations of 3-mm Polypropylene beads in a bed of small Glass beads: ({\Large{$\circ$}}) 50 beads (corresponding to 3\% volume fraction), ({\large{$\times$}}) 17 beads (1\%). {\bf For Figs.~\ref{r/Rpropyl}-\ref{r/Rplomb}, \ref{r/Rdensity1}-\ref{zirblast} standard deviations of the $r_i$ distributions are represented as errors bars. All beads touching the periphery corresponds to $r/R$=$r/R_{max}$. Lines provided to illustrate data trends: \cite{fit} for Figs.~\ref{r/Rpropyl}-\ref{r/Rplomb}, \cite{fit2} for Figs~\ref{r/Rdensity1}-\ref{r/Rdensity}.} Here, $r/R_{max}$=0.938. Density ratio is 0.36.}
\label{r/Rpropyl}
\end{figure}

\begin{figure}[htpb]
\center
\includegraphics[width=7.4cm]{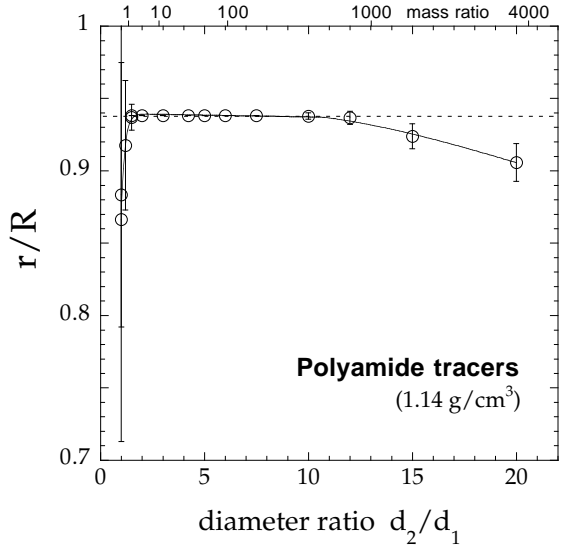}
\caption{Mean locations of 3-mm Polyamide beads in a bed of small Glass beads: ({\Large{$\circ$}}) 50 beads (3\%). See caption Fig.~\ref{r/Rpropyl}. $r/R_{max}$=0.938. Density ratio is 0.45.}
\label{r/Ramide}
\end{figure}

\begin{figure}[htpb]
\center
\includegraphics[width=7.4cm]{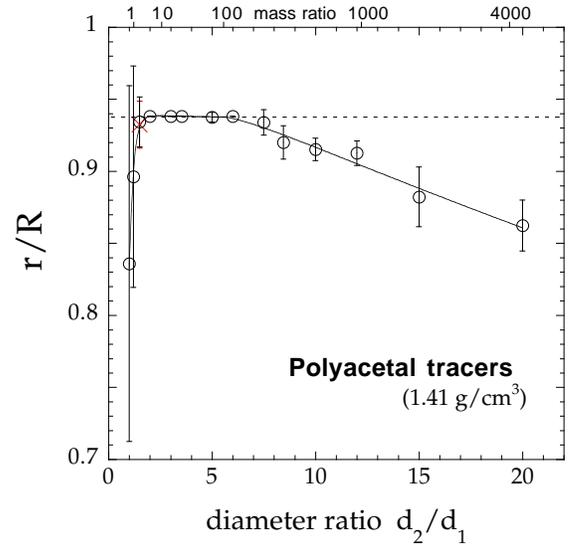}
\caption{Mean locations of 3-mm Polyacetal beads in a bed of small Glass beads: ({\Large{$\circ$}}) 50 beads (3\%), ({\large{$\times$}}) 17 beads (1\%). See caption Fig.~\ref{r/Rpropyl}.  $r/R_{max}$=0.938. Experiment with size ratio 3.5 has been done at 0.15~s$^{-1}$. Density ratio is 0.56.}
\label{r/Racetal}
\end{figure}

\begin{figure}[htpb]
\center
\includegraphics[width=7.4cm]{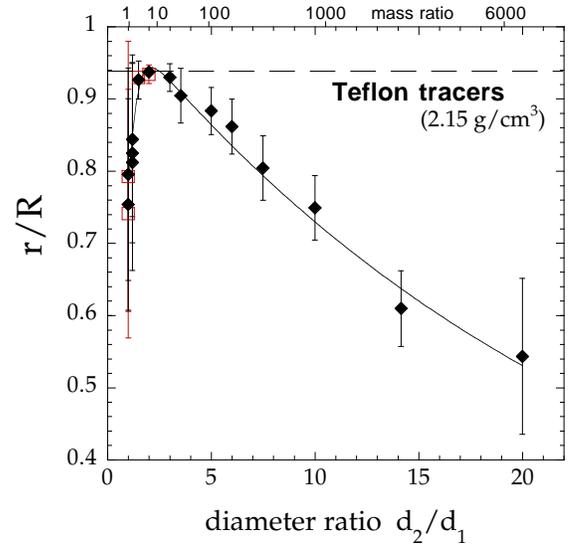}
\caption{Mean locations of 3-mm Teflon beads in a bed of small Glass beads: ($\blacklozenge$) 50 beads (3\%), ({\footnotesize{$\square$}}) 17 beads (1\%). See caption Fig.~\ref{r/Rpropyl}. $r/R_{max}$=0.938. There are strong electrostatic effects for size ratio 20. Density ratio is 0.86.}
\label{r/Rteflon}
\end{figure}

\begin{figure}[htpb]
\center
\includegraphics[width=7.4cm]{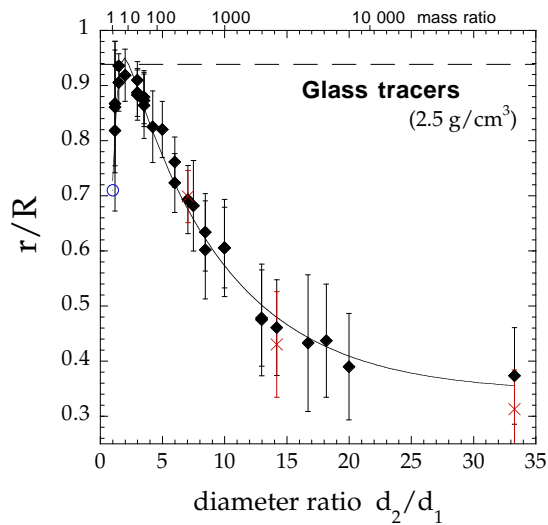}
\caption{Mean locations of 3-mm Glass beads in a bed of small Glass beads: ($\blacklozenge$) 50 beads (3\%), ({\large{$\times$}}) 17 beads (1\%), ({\Large{$\circ$}}) theoretical homogeneous point ($d_2$/$d_1$=1; $r/R$=0.71). See caption Fig.~\ref{r/Rpropyl}. $r/R_{max}$=0.938. Compared to Fig.~14 presented in \cite{moi}, some results have been added, and some have been slightly corrected by removing the tracers touching the caps when it was possible. Standard deviations are larger than for other materials due to the less precise hand of the experimenter at that time. The experiment with size ratio 7 (3\%) and the upper result for size ratio 6 have been done at larger speeds, resp. 0.11 and 0.13~s$^{-1}$. Density ratio is 1.}
\label{r/Rverre}
\end{figure}

\begin{figure}[htpb]
\center
\includegraphics[width=7.4cm]{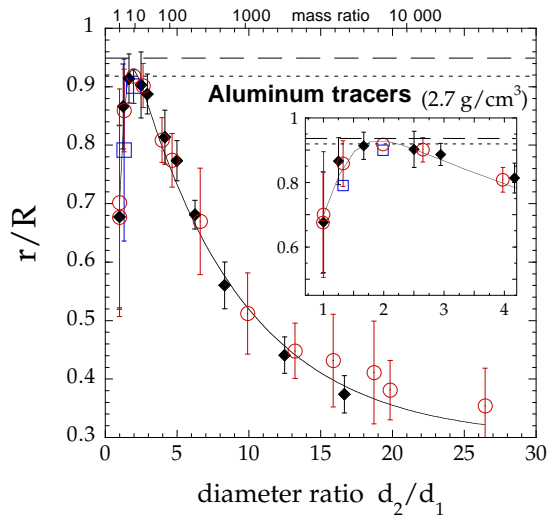}
\caption{Mean locations of Aluminum beads in a bed of small Glass beads.  3.97-mm tracers: ({\Large{$\circ$}}) 23 beads (3\%) and ({\footnotesize{$\square$}}) 15 beads (2\%);  2.5-mm tracers: ($\blacklozenge$) 20 beads (0.65\%). See caption Fig.~\ref{r/Rpropyl}. $r/R_{max}$=0.918 for large tracers, and to 0.948 for small tracers. The line takes into account both sets of data. Density ratio is 1.08.}
\label{r/Ralu}
\end{figure}

\begin{figure}[htpb]
\center
\includegraphics[width=7.4cm]{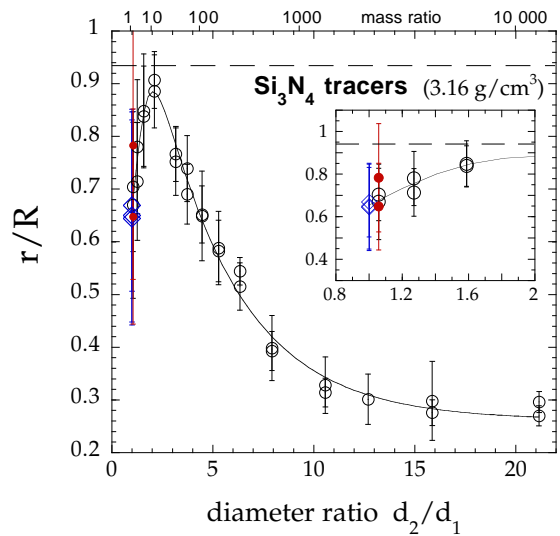}
\caption{Mean locations of Si$_3$N$_4$ beads in a bed of small Glass beads: ({\Large{$\circ$}})
10 beads (0.7\%) of 3.175~mm,  ({$\lozenge$}) 10 beads (0.6\%) of 3~mm (only for size ratio 1), ($\bullet$) avalanches regime for size ratio 1.06. See caption Fig.~\ref{r/Rpropyl}. $r/R_{max}$=0.934. Each experiment has been done twice, except for size ratios 12.7 (once) and 1 (3 times). Density ratio is 1.26.}
\label{r/RSi3N4}
\end{figure}

\begin{figure}[htpb]
\center
\includegraphics[width=7.4cm]{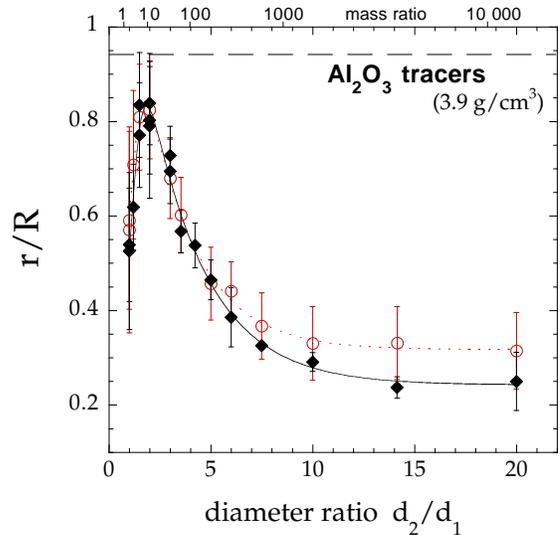}
\caption{Mean locations of 3-mm Al$_2$O$_3$ beads in a bed of small Glass beads: ({\Large{$\circ$}}) 50 beads (3\%), ($\blacklozenge$) 10 beads (0.6\%). See caption Fig.~\ref{r/Rpropyl}. $r/R_{max}$=0.938. The line for 0.6\% is used in Fig.~\ref{lesfitr/R}. Density ratio is 1.56.}
\label{r/Ral2o3}
\end{figure}

\begin{figure}[htpb]
\center
\includegraphics[width=7.4cm]{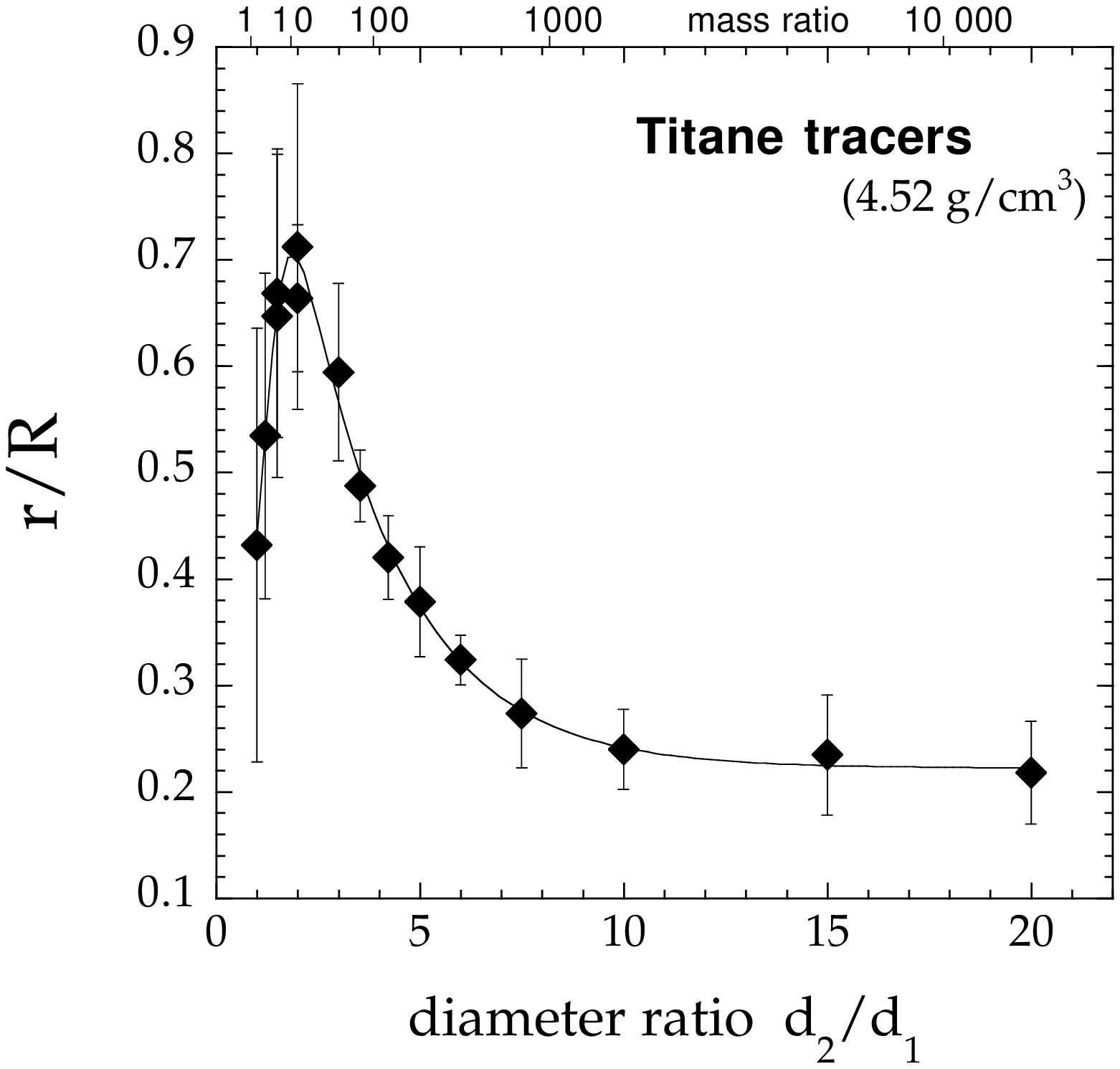}
\caption{Mean locations of 3-mm Titane beads in a bed of small Glass beads: ($\blacklozenge$) 10 beads (0.6\%). See caption Fig. \ref{r/Rpropyl}. Density ratio is 1.8.}
\label{r/RTitane}
\end{figure}

\begin{figure}[htpb]
\center
\includegraphics[width=7.4cm]{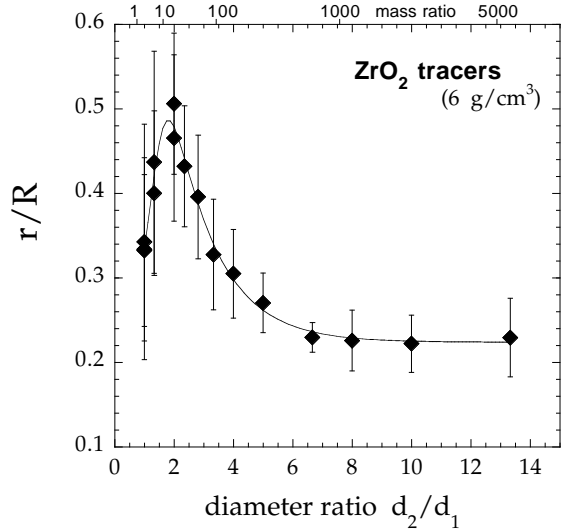}
\caption{Mean locations of 2-mm ZrO$_2$ beads in a bed of small Glass beads: ($\blacklozenge$) 25 beads (0.45\%). See caption Fig. \ref{r/Rpropyl}. Experiments with 1.3 and 2 size ratios have been done at 0.15~s$^{-1}$. Density ratio is 2.4.}
\label{r/RZrO2}
\end{figure}

\begin{figure}[htpb]
\center
\includegraphics[width=7.4cm]{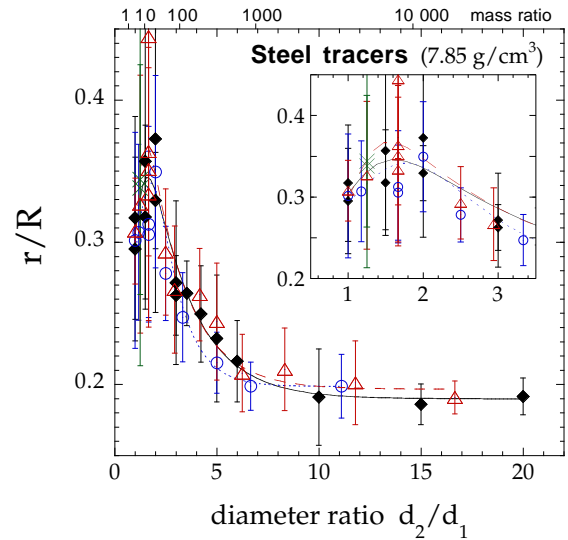}
\caption{Mean locations of Steel beads in a bed of small Glass beads: ($\blacklozenge$ and solid line (used in Fig.~\ref{lesfitr/R})) 10 beads of 3~mm (0.6\%), ($\triangle$ and dashed line) 15 beads of 2.5~mm (0.47\%), ({\Large{$\circ$}} and dotted line) 72 beads of 1~mm (0.16\%). See caption Fig.~\ref{r/Rpropyl}. For 2.5-mm beads, experiments with size ratios 8.3 and 11.8 have been done at a larger speed (0.15~s$^{-1}$) and ({\large{$\times$}}) corresponds to avalanches regime. Density ratio is 3.1.}
\label{r/Racier}
\end{figure}

\begin{figure}[htpb]
\center
\includegraphics[width=7.4cm]{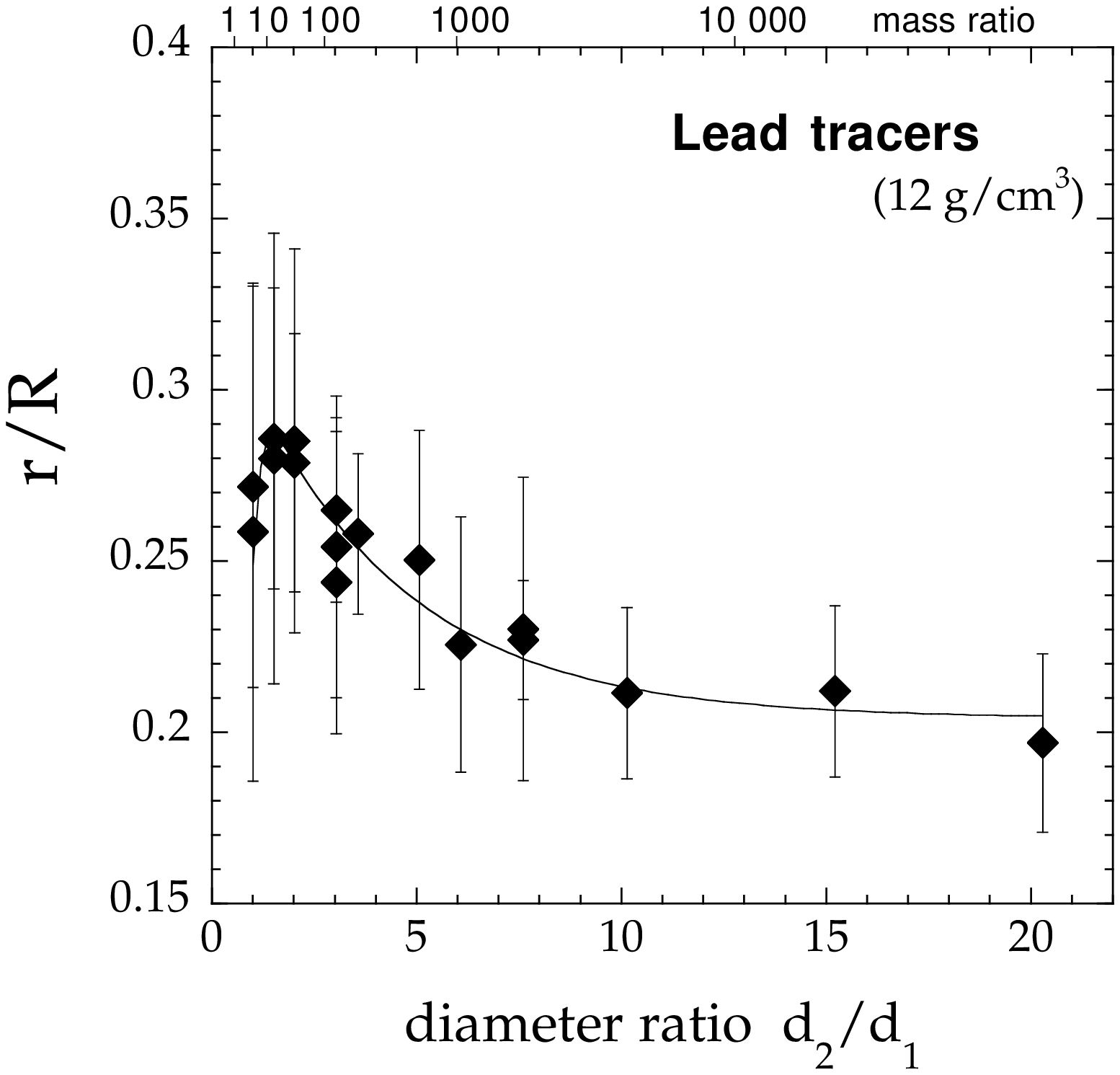}
\caption{Mean locations of 3-mm Lead beads in a bed of small Glass beads: ($\blacklozenge$) 10 beads (0.6\%). See caption Fig.~\ref{r/Rpropyl}. Density ratio is 4.8.}
\label{r/Rplomb}
\end{figure}

\begin{figure}[htpb]
\center
\includegraphics[width=8.6cm]{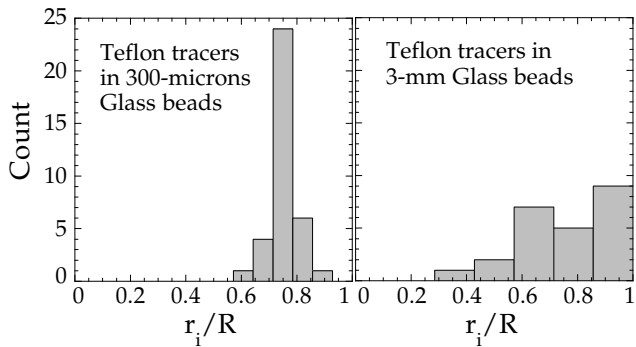}
\caption{Histograms of locations of 3-mm Teflon beads in a bed of Glass beads. Both mean values $r/R$ are similar but correspond (left) or not (right) to the most probable value.}
\label{histo}
\end{figure}

Roughly, for each experiment, excepting when $d_2/d_1$$\leqslant$2, the standard deviations of the $r_i$ distributions [Figs.~\ref{r/Rpropyl}-\ref{r/Rplomb}] are small: there is a well-defined segregation pattern. We call it a strong segregation process. In that case, $r$, the average of the $r_i$ distribution, is equal, or very close, to the position where one tracer would preferentially be. In other words, the mean value, $r$ is equal to the maximum of the $r_i$ distribution~[Fig.~\ref{histo}].

\begin{figure}[htpb]
\center
\includegraphics[width=3.6cm]{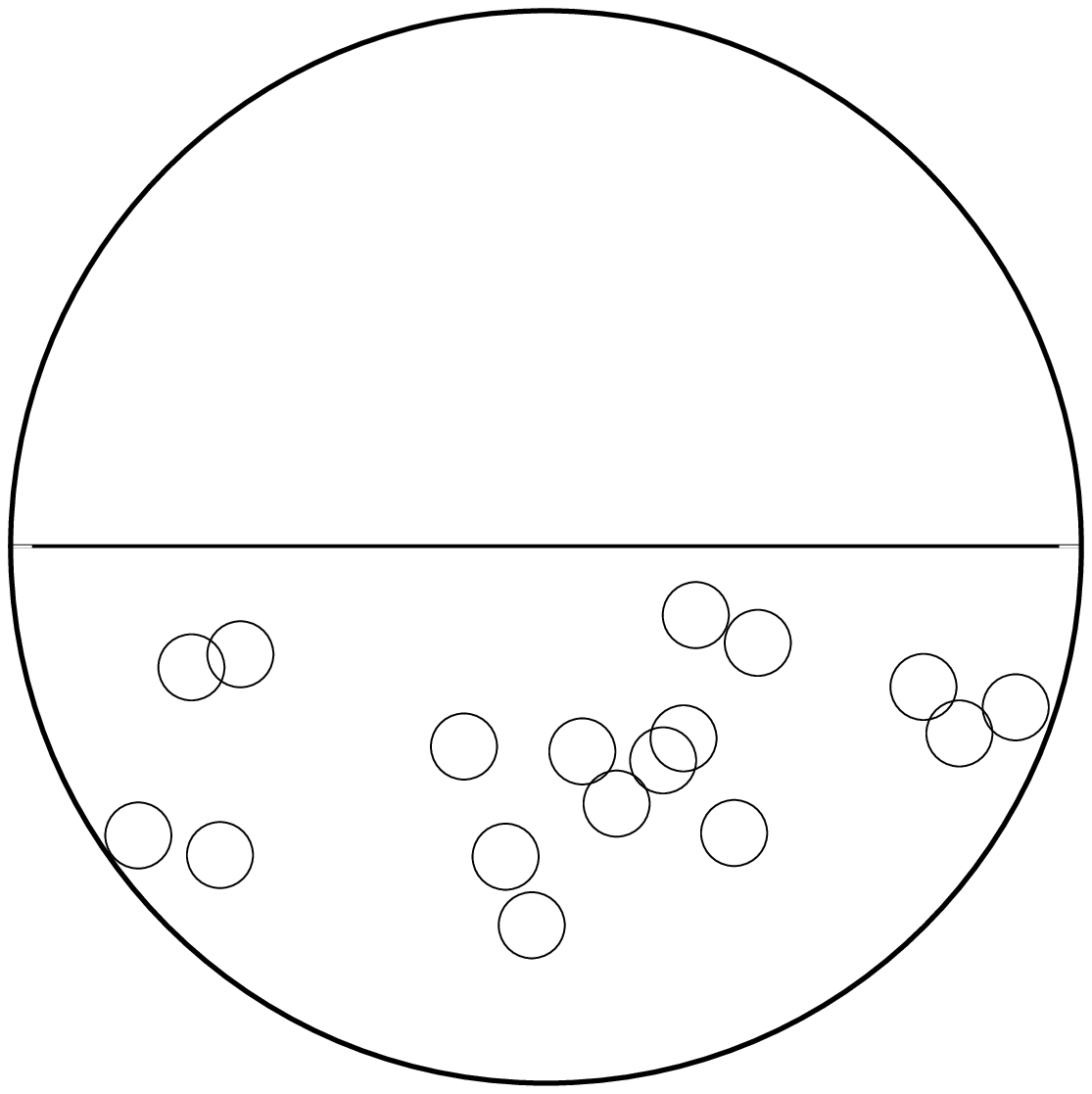}
\caption{3-mm Si$_3$N$_4$ tracers in a bed of 3-mm Glass beads.}
\label{nonseg}
\end{figure}

For the smallest size ratios, and especially for $d_2/d_1$$=$1, the standard deviations are large: the segregation pattern is not well-defined and the beads are spread [Fig.~\ref{nonseg}]. We qualify the segregation process as weak. $r$ can be significantly different from the maximum probability value, when the distribution is non symmetrical~[Fig.~\ref{histo}].

Here, we discuss what information could be deduced on the segregation effects, coming from observations made on the focus of the segregation pattern.
Tracers have been driven to their final position due to some forces exerted on them. We could separate them into forces due to the size difference (geometrical force), due to the mass or density difference (buoyancy, weight), due to a shape difference, {\it etc}. The sum of all these forces determines the final location. Two cases have to be distinguished to interpret the focus of the segregation pattern. (1) When beads are at the top or at the bottom, the total force resulting of all {\it segregation} effects is not equal to zero. It is in balance with an iceberg effect near the free surface (the weight of the emerging part of the tracer), or with the effect of the static bed at the bottom of the flow (floor effect). The focus of the pattern depends on the intensity of the total segregation force. (2) When beads are at an intermediate level, the segregation effects balance at that level, and the total segregation force is equal to zero at that level. In that case, the dependency of the total force with the position $z$ determines the focus of the pattern. When the total segregation force strongly depends on $z$, the pattern is well-focused around the balance level.

In both cases, the intensity of the force or of its $z$-derivative has to be compared with the fluctuations of the tracer trajectories.
We expect that the amplitude of the fluctuations depends on $d_2$/$d_1$, and on the flow thickness in number of small beads, related to $d_1/R$.
Fluctuations are also enhanced when tracers interact.\\

Roughly, results show three tendencies: the focus is very strong for tracers at the periphery, strong for high size ratios and weak for small size ratios. Very strong focus when tracers saturate at the periphery indicates that a non zero total segregation force is an efficient effect of segregation. This is not observed when tracers saturate at the center because it is masked by the interaction between tracers which creates fluctuations. The use of a larger amount of tracers (3\% instead of 0.6\%) [Fig.~\ref{r/Ral2o3}] increases the standard deviations when $r$ tends to 0.

Several causes can explain the weak segregation at small diameter ratios. It could be due to a decrease of the segregation force or its $z$-derivative, or due to the enhancement of the fluctuations of trajectories. Fluctuations enhancement could be characteristic of small size ratios or could come from the fact that excessively large beads may have been used in the device. In fact, in our experiments, small diameter ratios are obtained by increasing the size of the small beads, which leads to beds composed of relatively large beads (4 to 1~mm) compared to the drum diameter (48.5~mm). The flowing layer is then only 12 to 48 beads long, which may not be enough to ensure a segregation process fully developed. Moreover, at low tracers fractions, the flow thickness is imposed by the small beads size \cite{gwenthese}: for such a geometry, the flow thickness is around 6 beads for 1-mm beads, and 1-2 beads for 3-mm beads. Such layers might not be considered as a continuous flowing material, but as a rolling layer of beads. In that case, trajectories of tracers and beads of the bed should be very chaotic and fluctuations are probably as large as the flow thickness. 

To differentiate between the two possibilities explaining possible enhanced fluctuations, we compare 3 (resp. 2) sets of experiments with Steel (resp. Aluminum) tracers of different sizes [Figs.~\ref{r/Racier}, \ref{r/Ralu}]. For the smallest tracers, we found that the increase of standard deviations at small size ratios still exists, although it concerns smaller sizes for the bed beads and thicker/longer flows measured in bed bead diameters. Moreover, in other devices (Sec.~\ref{recursive}), some mixtures with a small size ratio look homogeneous, whatever is the size of the bed particles compared to the flow thickness. These two points prove that the weak segregation is not due to the use of excessively large beads for the bed, compared to the drum size. In conclusion, the segregation pattern is focused at large size ratios, and is not well-defined for small size ratios, because of the variations of either the segregation force or the trajectory fluctuations. This point is independent of the value of the density ratio.
 
Consequently, density and size ratios do not play an equivalent role in determining the focus of the pattern. Considering tracers {\it at intermediate location}, we deduce that the segregation force due to a density ratio weakly depends on $z$ (large standard deviations for $d_2/d_1$=1). But, the data do not give deductions about the gradient of the segregation force coming from the geometrical effect: small values of the standard deviations can be due to small trajectories fluctuations or to a strong $z$-dependency of the geometrical force. Other arguments based on $r/R$ data (end of Sec.~\ref{grad}) show that there is a $z$-gradient of the geometrical force.\\

We also would like to point out a possible statistical effect due to spread asymmetric patterns [Fig.~\ref{histo}], which could influence the interpretation of mean locations $r/R$ in terms of balance level.
Although $r/R$ values for the weak segregations ($d_2/d_1$$<$1.5) vary regularly with the density ratio [Fig.~\ref{r/Rdensity1}] and are consistent with values obtained for larger size ratios [Figs.~\ref{r/Rpropyl}-\ref{r/Rplomb}], $r/R$ could be significantly different from the maximal probability of $r_i$. 
Assuming two preferential positions, the top for light tracers and the bottom for dense tracers, the variation of $r/R$ with the density ratio could be just a consequence of a variable spreading of the distribution: the closer to 1 the density ratio is, the wider the distribution could be. The available set of data is not statistically large enough to differentiate between a variation of the preferential location and those of the spreading of the distribution. 

In conclusion, we keep in mind that first values of the $r/R$ vs $d_2/d_1$ curves may have to be shifted to 1 (resp. 0) for materials lighter (resp. denser) than Glass to obtain the curves of the preferential locations (balance levels). We also know that the $r/R$ curves are  almost equal to those of the preferential locations when standard deviations are small, which is the case of most experiments.

\subsubsection{Location of the tracers for each material}\label{cloche}

In these experiments, we vary the size of the small beads and the density of the tracers. We also change the density of the small beads, to check that the ring position depends on the density ratio rather than on density (Sec.~\ref{grad}). Tracers location has been shown to depend on the size ratio rather than on the sizes \cite{moi}. Consequently, we will use density and size ratios as the parameters controlling the ring position. \\

For each material, $r/R$ presents the same evolution with $d_2/d_1$: (1) a rapid increase, then (2) a long decrease [Figs.~\ref{r/Rpropyl}-\ref{r/Rplomb}]. The increasing part is restricted to small size ratios, roughly $<$2, giving a maximum for $r/R$ corresponding to size ratios between 1.5 and 2. 
We will see (Sec.~\ref{recursive}) that indications of this complex variation of $r/R$ vs $d_2/d_1$, are found in other experimental devices.\newline
(1) The increase corresponds to the classical up-segregation of large beads, intensively studied by the granular community. The `up-character' is stronger when the size ratio increases. When $d_2/d_1$ passes from 1 to 2, $r/R$ increases, and reaches the saturation value for low densities: all tracers touch the drum periphery. For larger densities, the maxium radius also occurs for size ratio close to 2, but the ring is buried inside the drum. \newline
(2) The long decrease corresponds to tracers getting larger, and consequently heavier. They progressively sink into the bed by a push-away process. This new phenomenon has been named intermediate and reverse segregations by the authors \cite{moi}. It happens even for tracers less dense than the Glass beads of the bed. For the dense materials, the ring radius tends to saturate to its minimum value related to the flow thickness.

\begin{figure}[htpb]
\center
\includegraphics[width=7.8cm]{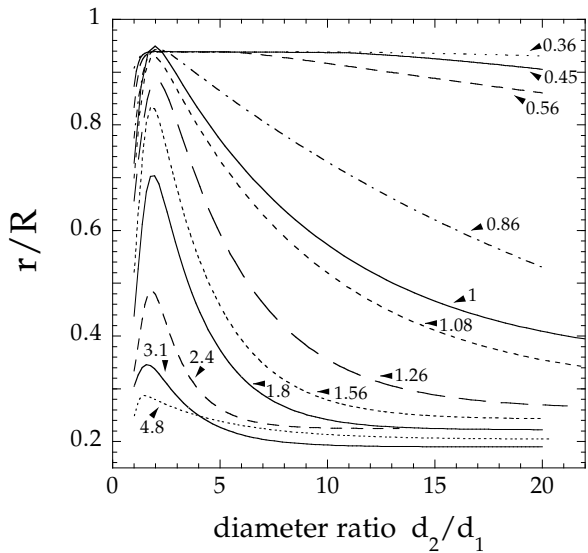}
\caption{Curves of mean locations $r/R$ of large tracers of different densities plunged in a bed of small Glass beads (from Figs.~\ref{r/Rpropyl}-\ref{r/Rplomb}). Density ratio $\rho_2/\rho_1$ is written on each curve.}
\label{lesfitr/R}
\end{figure}

The evolutions are similar to those of the location of a large Glass tracer in a bed of small Glass beads, corresponding to $\rho_2/\rho_1$=1 (\cite{moi} and Fig.~\ref{r/Rverre}). The theoretical point for $d_2$/$d_1$=1, {\it i.e.} the homogeneous media, would correspond to $r/R$=2/3 in a half-filled drum, but because we remove all tracers which are closer to the free surface than 0.18$R$, a numerical calculation gives $r/R$=0.71 for homogeneity.
Compared to those of Glass tracers, the $r/R$ curves are shifted to the top for lighter materials, or to the bottom for denser materials (note on Lead in \cite{lead}) [Fig.~\ref{lesfitr/R}]: both evolutions with $d_2/d_1$ and $\rho_2/\rho_1$ are continuous.
The continuity could appear surprising and we could have expected tracers denser (resp. lighter) than Glass at the bottom (resp. at the top) of the flow. The balance at intermediate level is not explained by buoyancy variations either (Sec.~\ref{grad}). Both density and size ratios take a role in the selection of the tracer location, as already found \cite{roseman}.\newline 

From each $r/R$ curve, we want to deduce the evolution of the preferential level, where all segregation forces balance. Both variables are linked when standard deviations are small ($d_2/d_1$$>$2) but could be different for large standard deviations ($d_2/d_1$$<$2).\newline
- For materials lighter than Glass ($\rho_2/\rho_1$$<$1), the most probable value of the $r_i$ distribution could be the periphery for $d_2/d_1$$\lesssim$2. This suggests that the $r/R$ increase could be a consequence of an asymmetric spreading of the $r_i$ distribution and not of a variation of the preferential location. In that hypothesis, the preferential location would be constant up to $d_2/d_1$$\approx$2, then decreases in the same way as $r/R$. \newline
- For denser materials ($\rho_2/\rho_1$$>$1), the expected most probable value is close to the center of the drum, which would enhance the increasing part of the curve. Consequently, asymmetry of histograms is not the reason for the $r/R$ increase. The evolution of the preferential level really follows a complex curve, increasing then decreasing, similarly to those of~$r/R$. \newline

One consequence of this complex variation is that two mixtures having different $d_2/d_1$ can show the same mean segregation pattern (same $r/R$), even if composed of the same materials. Two sizes of tracers will then be found at the same place in a bed of an other material beads (for example Si$_3$N$_4$ beads with 1.4 and 3 size ratios in small Glass beads [Fig.~\ref{r/RSi3N4}]). This gives possibilities of mixing between two minority species, with the same or with different densities, inside a third majority component.

\subsubsection{Influence of the bead density on the tracer location}\label{grad}

\begin{figure}[htpb]
\center
\includegraphics[width=7.8cm]{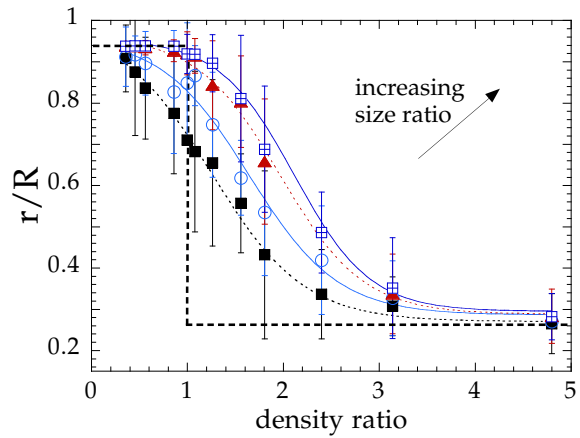}
\caption{Tracers mean locations for size ratios: ({\small{$\blacksquare$}}) 1 (and few ratios comprised in 0.992-1.015), ({\Large{$\circ$}}) 1.2 (1.25-1.33), ({\large{$\blacktriangle$}}) 1.5 (1.52-1.67), ($\boxplus$) 2 (1.98-2.11). See caption Fig.~\ref{r/Rpropyl}. 
The dashed step line is one possible curve for the preferential location for $d_2/d_1$$=$1 (see Sec.~\ref{artefact}).}
\label{r/Rdensity1}
\end{figure}

\begin{figure}[htpb]
\center
\includegraphics[width=7.8cm]{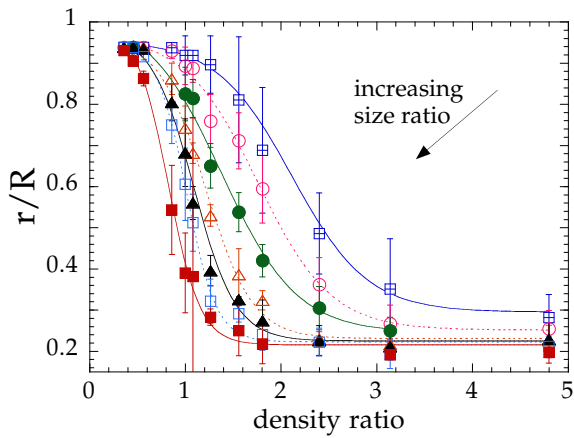}
\caption{Tracers mean locations for size ratios: ($\boxplus$) 2 (and few ratios comprised in 1.98-2.11), ({\Large{$\circ$}}) 3 (2.94-3.175), ({\Large{$\bullet$}}) 4.23 (4-4.47), ($\triangle$) 6 (6-6.67), ({\large{$\blacktriangle$}}) 7.5 (7.5-8.33), ({\small{$\square$}}) 10 (9.92-10.58), ({\small{$\blacksquare$}}) 20 (19.85-21.17) (averaged data when there are several experiments). See caption Fig.~\ref{r/Rpropyl}.}
\label{r/Rdensity}
\end{figure}

Tracers denser or lighter than Glass generally reach an intermediate level. For a given size ratio, the denser the tracer is, the deeper is its position [Figs.~\ref{r/Rpropyl}-\ref{r/Rplomb}]. We can continuously vary its position keeping the geometry constant (same $d_2/d_1$, same geometrical effect) [Figs.~\ref{r/Rdensity1},~\ref{r/Rdensity}]. 

When increasing $\rho_2/\rho_1$, a large bead passes from the surface to the bottom. This sinking appears earlier for large size ratios than for small ones because large tracers are already heavy [Fig.~\ref{r/Rdensity}].
But for size ratios 1 to 2 [Fig.~\ref{r/Rdensity1}], the sinking is delayed when $d_2/d_1$ is large. It corresponds to the sharp increase of the $r/R$ vs $d_2/d_1$ curves [Figs.~\ref{r/Rpropyl}-\ref{r/Rplomb}]. For $d_2/d_1$=1, we could have expected $r/R$ maximum for all lighter tracers than Glass, and $r/R$ minimum for all denser tracers (step line in Fig.~\ref{r/Rdensity1}). But this is not what it is observed experimentally [this study] and numerically \cite{ristow94}. $r/R$ varies continuously, and suggests that a pure difference of density leads to a balance at an intermediate level in the flow. Quantitatively, our curve decreases more rapidly and goes further down than the 2D simulations \cite{ristow94}, probably because of a reduced ability for the tracer to move down in 2D. Nevertheless, we have to take into account statistical effects (Sec.~\ref{artefact}) to determine the curve of the preferential location, which is probably between the step line and the $r/R$ curve [Fig.~\ref{r/Rdensity1}]. Further investigations have been planned to determine if the balance at an intermediate level due to a pure difference of density experimentally happens or not. \\

\begin{figure}[htpb]
\center
\includegraphics[width=7.8cm]{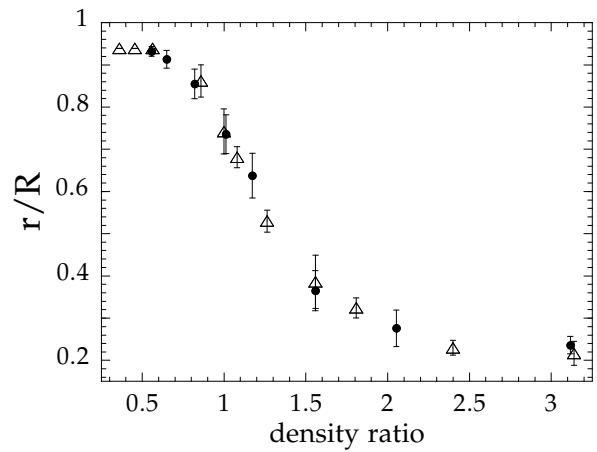}
\caption{Mean locations of 3-mm tracers: ($\triangle$) in 2.5-g/cm$^3$ Glass beads (data from Fig. \ref{r/Rdensity}), ({\large{$\bullet$}}) in 3.85-g/cm$^3$ Zirblast beads. In both cases, size ratio is 6. See caption Fig.~\ref{r/Rpropyl}.} 
\label{zirblast}
\end{figure}

Additional experiments involving a bed of small Zirblast beads instead of small Glass beads have been done. Zirblast beads are 500~$\mu$m in diameter and 3.85~g/cm$^3$ in density, mainly composed of ZrO$_2$ (67\%) and SiO$_2$ (30\%). Each experiment has been done at 0.085~s$^{-1}$, with 10 tracers. Tracers are 3-mm beads made of Teflon, Glass, Si$_3$N$_4$, Al$_2$O$_3$, Titane, Zircone, Steel (7.91~g/cm$^3$) and Lead (Table~\ref{billes}). The positions of tracers for experiments done with small Glass and those done with small Zirblast beads plot very well on the same curve [Fig.~\ref{zirblast}]. We deduce that $r/R$ depends on the density ratio rather than on the density. \newline

The density difference of the material is not the only reason for a large bead to be buried in the flowing layer. It could not explain the fact that large Glass beads or large beads made of lighter materials can be buried [Fig.~\ref{r/Rteflon}]. These observations prove that there is a mass effect involved in the segregation process: the large tracers are heavy enough to push away their small neighbors and to go down. Thinking in terms of Fluid Mechanics, the mass effect can be interpreted by the fact that a Glass tracer is denser than a bed of small Glass beads containing at least 27\% of void. Consequently, the density of the tracer has to be compared to the one of a small bead for $d_2/d_1$$\approx$1, or to the density of the bed for $d_2/d_1$$\gg$1. This interpretation implies that tracers lighter than the bed are always visible in surface, their partial sinking being due to an iceberg effect [Fig.~\ref{lesfitr/R}].\newline

Each tracer is affected by the mass and the geometrical effects and its position is determined by the addition of these effects. By varying the density ratio between the beads, only the mass effect changes. But, by varying the size ratio, both effects (geometrical and mass) change. When increasing the tracer size, the mass effect tends to be dominant explaining that larger tracers get buried deeper. We represent each effect by a force exerted on the tracer; the mass effect by the sum of the weight and the buoyancy, the geometrical effect by the result of the actions of the small beads on the tracer (except the buoyancy). In that point of view, the mass effect depends on density and size ratios, the geometrical effect only on the size ratio. The balance is obtained where the sum of the forces is equal to zero.

Considering a tracer at an intermediate level within the flow, the mass and the geometrical effects exerted on it balance: it shows no vertical relative motion. If the balance is obtained at that particular level $z_i$, it means that, at least, one of the two effects depends on the vertical coordinate $z$ within the flow. Hereafter, we discuss the possible dependencies of the two effects with~$z$.

Because the focus of the pattern is weak for $d_2$/$d_1$=1, we do not expect that the mass effect strongly depends on $z$. Nevertheless, the mass effect includes the buoyancy which varies with $z$ due to the variation of compacity inside the flow. Only a few studies present data on vertical gradient of compacity in the flowing layer of a drum \cite{Rajchenbach, bonamy, GDR, olivierchevoir}, but they concern 2D or quasi-2D experimental devices. Excepting the first layer of beads in which the compacity drops to zero,  2D compacity ranges from 0.4 to 0.8. From 3D numerical simulations of a drum flowing layer, an estimate of the maximal compacity variation is from 0.4 to 0.6 \cite{umberto}. Consequently, the maximal range of buoyancy is $g V$[0.4-0.6]$\times$2.5~g/cm$^3$ in small Glass beads, where $g$ is the gravity, and $V$ the volume of the tracer. 

The question is to estimate if this range of buoyancy is large enough to explain that tracers having different densities reach equilibrium at different levels inside the same flow. We limit the calculus to tracers totally embedded, but not touching the bottom of the flow: their density varies from 2.15 to 6~g/cm$^3$ (Sec.~\ref{cloche}), leading to a range of weights of $g V$[2.15-6]~g/cm$^3$, approximately 10 times bigger than the range of buoyancy.
Although there is a large uncertainty on the compacity, the variations of buoyancy are too small to balance the range of tracer weights. We can not explain our results by involving only a mass effect variation with~$z$.
Moreover, we found tracers {\it of the same size} and of different materials located at intermediate levels, {\it i.e.} with mass and geometrical effects in balance, although they induce very different values of the mass effect (very different weight, similar buoyancy). This implies that they are submitted to very different values of the geometrical effect. Because they have the same size, these different values are due to the fact that they are located at different levels. We conclude that the geometrical effect strongly depends on $z$, while the mass effect varies weakly with~$z$.\newline

The data also show that the density differences prevail for large and small density ratios [Figs.~\ref{r/Rdensity1},~\ref{r/Rdensity}]. Very light tracers ($\rho_2$/$\rho_1$$\lesssim$0.5) are always at the surface regardless of size and very dense tracers ($\rho_2$/$\rho_1$$\gtrsim$4) are always at the bottom. For those extreme density ratios, the preferential location of a tracer is not influenced by the value of its diameter. In these cases, 
there is no balance between the two effects: the mass effect always dominates for very dense beads, or is acting upwards for very light beads, as the geometrical effect does. While for all intermediate density ratios, the size ratio has an influence: the whole pair ($d_2/d_1$, $\rho_2/\rho_1$) determines the position of the bead. 

\subsubsection{A few more remarks}\label{rmq}

Two sets of experiments have been done with Si$_3$N$_4$ tracers [Fig.~\ref{r/RSi3N4}]. Both $r/R$, each calculated on the ten $r_i$ data, are similar, which is not surprising in view of the small standard deviations of each set. The two sets show the good reproducibility of the results.\newline

Experiments with the 3 Steel tracers (1, 2.5 and 3~mm) [Fig.~\ref{r/Racier}], and those with the 2 Aluminum tracers (2.5 and 4~mm) [Fig.~\ref{r/Ralu}] show that the size ratio is the relevant parameter when tracers are at intermediate level, as already found with Glass tracers \cite{moi}. Small effects due to the tracer size might appear when they are near center or  periphery, as the Aluminum biggest tracers up-shift close to the center of the drum.\newline

We do not notice any influence of the rotation speed, randomly distributed between 0.04 to 0.1~s$^{-1}$. It is probably because the flow thickness is weakly dependent of the rotation speed \cite{gwen} for that range of drum and beads sizes (geometrically, increasing the flow thickness slightly increases $r/R$, significantly only for small $r/R$). Both sets with Si$_3$N$_4$ tracers have been done at speeds around 0.085~s$^{-1}$ and 0.15~s$^{-1}$, showing no difference, and no systematic relative order between the two sets [Fig.~\ref{r/RSi3N4}]. 
Some of the experiments with the largest `small' Glass beads (2-3~mm), usually done at high speeds, have also been done at 0.085~s$^{-1}$ at which the flow is in the avalanches regime [Figs.~\ref{r/RSi3N4}, \ref{r/Racier}]. Tracers positions are more spread but $r/R$ are very close to those at higher speed, when averaged on several experiments.\newline

The influence of the volume fraction of large tracers on their final location has been previously tested on Glass tracers \cite{moi}: interactions between tracers shift the segregated locations towards the surface. The interpretation involves the reduction of the push-away process due to the presence of other large (and heavy) tracers in the bed. The shift is combined with a spreading of the positions: for $d_2/d_1$ such that a few tracers segregate at a deep intermediate level, a nearly homogeneous mixture is experimentally obtained by increasing the fraction of tracers. 
Consequently, two behaviors are expected when increasing the fraction of tracers in a mixture of beads of different densities and sizes: (1) for very large or dense large tracers the mixture tends to be more homogeneous; (2) for light large tracers, the mixture still segregates with large beads on the surface. These 2 evolutions are compatible with the results of numerical simulations \cite{dury99} obtained on 50-50\% mixtures, showing segregation for light large beads and homogeneous mixing for dense large beads. In the same way, the value of $r/R$ at low fraction can be used to estimate the segregation state of a mixture at high fractions. For example, in a 3D drum, 50\% of large `tracers' with a size ratio 3, and a density ratio 1.5, which would lead to $r/R$=0.7 in our device, segregate at the surface \cite{Hill}. In a 2D drum, a mixture with size ratio 1.14 (resp. 1.42), density ratio 0.82 (resp. 1.53) and thus $r/R$=0.83 (resp. 0.75) in our device, stays homogeneous at 50\%, while a mixture with size ratio 1.14 (resp. 1.18), density ratio 1.53 (resp. 0.69), and $r/R$=0.6 (resp. 0.9), gives down (resp. up) segregation of large particles~\cite{metcalfe}.  

In our experiments, we keep the volume fraction low (from 0.16 to 3\%) and quantify its effect in Al$_2$O$_3$ experiments (3\% and 0.6\%) [Fig.~\ref{r/Ral2o3}] and on a few other data [Figs.~\ref{r/Rpropyl}, \ref{r/Racetal}-\ref{r/Ralu}]. 
Results show that there is almost no influence of the fraction in this small range. Nevertheless, when tracers are close to the center ($r/R$ lower than $\sim$0.4), the value of $r/R$ is slightly larger when using more tracers [Fig.~\ref{r/Ral2o3}]. A similar effect might be seen on Fig.~\ref{r/Ralu}, again for $r/R$ getting close to the center. The shift is small, but strong enough to make two `parallel' curves with different tracers fractions cross each other when getting to small $r/R$ values (high size ratios in Si$_3$N$_4$-0.7\% and Al$_2$O$_3$-3\% data [Figs.~\ref{r/RSi3N4}, \ref{r/Ral2o3}]). The $r/R$ up-shift comes with an increase of the standard deviations. It probably happens because too many tracers can not be all grouped in the center.
Moreover, the interaction between tracers enhances the trajectories fluctuations. It could prevent them from reaching the deeper point of the flowing layer because tracers roll on each other (one being in the flow, one in the solid rotating part). For that reason, only few tracers have been used in the case of Steel and Lead [Figs.~\ref{r/Racier}, \ref{r/Rplomb}]. For an obstruction reason too, we expected $r/R$ slightly smaller for larger fractions, when all beads are close to the periphery. But, this down-shift would appear for higher differences of fraction due to the larger volume available at the periphery than at the center, and we have not observed it.  In conclusion, for small fractions ($<$3\%), $r/R$ variations with the fraction are negligible, except when tracers are all grouped at the center.

\section{Diagram of surface segregation} \label{recursive}

\begin{figure}[htbp]
\center
\includegraphics[width= 4.05cm]{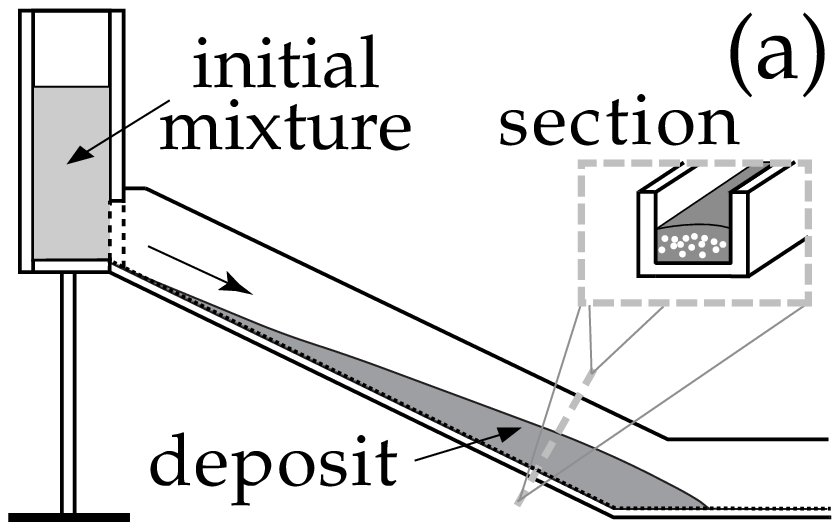}
\includegraphics[width=1.55cm]{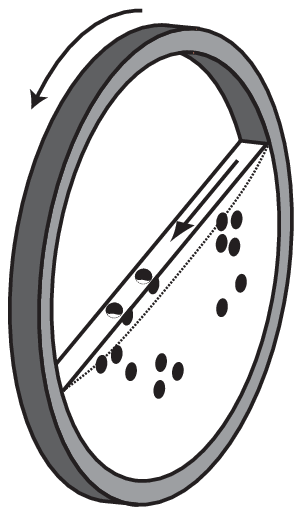}
\includegraphics[width=2.85cm]{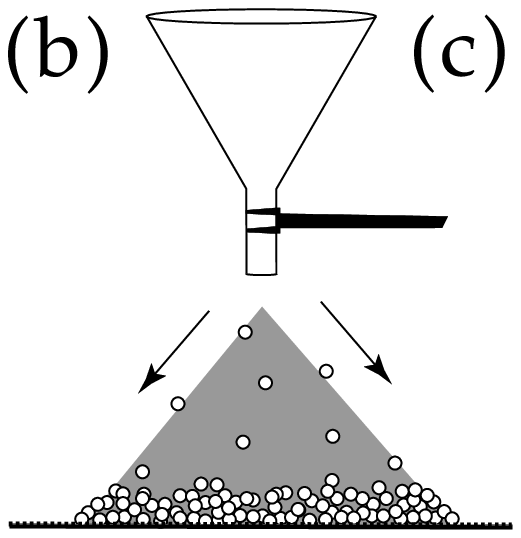}
\caption{Experimental devices for: (a) flows in a channel, (b) rotating a half-filled quasi 2D disc, (c) building a pile.}    
\label{devices}
\end{figure}

The complex shape of the $r/R$ vs $d_2/d_1$ curves indicates that the same mean position can be reached for two diameter ratios, even with tracers of the same density (Sec.~\ref{cloche}). To determine if that feature only happens in the drum device, or if it might be due the flows only a few beads thick, we work with 3 other experimental devices in which the flow thickness varies:  a channel, a half-filled quasi 2D disc (no bulk observations), and build-up of a pile [Fig.~\ref{devices}]. We explore the possibility for different tracers to be at the same level in the flow. 

For practical reasons, we choose as a criterion the presence or absence of tracers at the surface. ``Visible'' corresponds to tracers being over the surface or partially embedded in the small particles. ``Hidden'' corresponds to tracers buried inside the bed, whatever their position, intermediate or at the bottom. On the limit separating ``visible" and ``hidden" states, the mixture looks as if there is no segregation: tracers do not appear to be preferentially at one location, neither inside, nor at the surface. But, we do not distinguish between real homogeneous mixtures and a bad-focused segregation at a level close to the surface, called `no apparent segregation'. It is sometimes possible to differentiate
the 2 cases by a cross-section of the deposit, but most of the time, experiments are
not accurate enough.   
From channel experiments done with large and small Glass beads \cite{moi}, the limit occurs for size ratio around 4.2, which corresponds to tracers at $r/R$=0.8 in the 3D drum. Taking into account the uncertainties, this is not far from $r/R$=0.71 of a homogeneous medium (Sec.~\ref{cloche}). By extension, we could assume that both really or apparently homogeneous cases are obtained for mixtures in which tracers are around $r/R$=0.7-0.8 in the drum (hypothesis shown true in Sec.~\ref{RC}). The limit ``visible-hidden" would then be close to the iso-position curves $r/R$=0.7-0.8.

In the following sections, we draw in ($\rho_2/\rho_1$; $d_2/d_1$) coordinates the two fields (hidden, visible). Then, we examine if the shape of the limit between the fields explains that different tracers of the same density can segregate at the same level. 

\subsection{Experimental devices}

In these experiments, mixtures are composed of tracer particles indexed 2 in a bed of particles indexed 1. Volume fractions are 2\% for disc and piles, 10\% for channel. Most of beads used previously are not adapted here, for size (disc) or available quantity (piles, channel) reasons. Large and small particles are either beads (preferentially used) or angular particles (Table~\ref{particules}). Angularity of particles is probably one reason for the limit not to be sharp (Sec.~\ref{chenal}). Diameters are sizes or mean sizes of ranges, defined by a sieving interval (Table~\ref{particules}). There is no systematic use of one type of particles to play the role of tracers, or bed. The initial mixture homogeneity is obtained by putting successive thin layers of both components in the starting box of the channel, in the funnel, or by putting tracers by hand all over the 2D drum. Some additional experiments with a initially de-mixed state, in both possible configurations, show: no major difference for the channel, a slightly longer convergence time to the same final state for the drum, a strong modification of the results for piles. The influence of the initial homogeneity mainly depends on the possibility of global mixing during the flow.\newline

\begin{table}
\caption{Particle densities and sizes.}
\begin{center}
\begin {tabular}{|l|c|l|}
\hline
  ~~ \underline{Materials :}& \underline{density}&\underline{sizes or ranges of 
sizes  }\\
& \small{(g/cm$^3$)}& (in $\mu$m if `mm' not written)\\
Glass beads&2.5&45-90; 70-110; 150-212; \\
 & &180-212; 150-180; 150-250; \\
&&300-400; 300-425; 425-600; \\
& &600-710; 600-800; 600-850; \\
& &710-800; {\small and} 90; 150; 180; \\
&&200; 212; 250; 300; 355; 400; \\
&&425; 500; 600; 710; 800; 850; \\
 & & in mm: 1.5; 2; 3; 4; 5; 7.5\\
Sugar balls&1.6&in mm: 0.85-1; 1-1.18; 1.4-2\\
Steel beads&7.78&4.7 mm\\
Diakon particles &1.14&106-212; 300-400; 300-425;\\
(ovoid)& & 600-850; 625-850\\
Silicone Carbide&3.2&150-212; 212-300; 300-425;\\
fragments (SiC)& &400-500; 500-600; 600-850;\\
& &850-1180; {\small and} 600\\ 
Wood Nut  &1.35&300-355; 450-560; 450-800;\\  
fragments &&630-710; 600-800; 710-800; \\
&&1-1.7 mm \\
Sugar fragments&1.6&150-250; 150-300; 300-425;\\
&&600-850; 850-1000; 1-2 mm\\
Sand&2.6&600-850\\
Glass fragments&2.5&300-425; 425-500\\
\hline
\end{tabular}
\label{particules}
\end{center}

\end{table}

\subsubsection{Rotation in a quasi 2D half-filled disc}
\begin{figure}[htpb]
\center
\includegraphics[width=7.8cm]{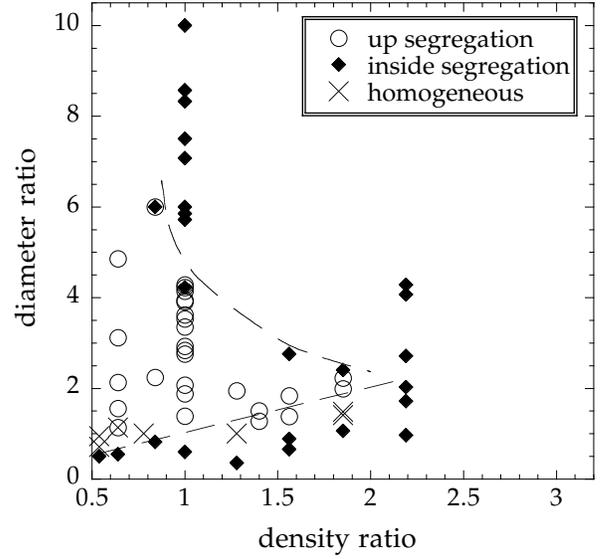}
\caption{Surface segregation in the {\bf disc}: tracers ({\Large{$\circ$}}) at the surface of the flow, ($\blacklozenge$) totally embedded or at the bottom, ({\large $\times$}) mainly embedded but still visible at the surface or homogeneous. {\bf In Figs.~\ref{RCdisc}-\ref{iso}, diameter and density ratios are for tracers to bed particles. Two symbols added means that there is no clear determination between these cases or that two experiments have been done. The same limit is drawn on Figs.~\ref{RCdisc}, \ref{RCtas}, \ref{RCchenal}, \ref{iso}.}}
\label{RCdisc}
\end{figure}

The drum is 3.5~mm thick, 81~mm in diameter, made of clear plastic [Fig.~\ref{anneau}]. The system is close to be 2D for largest tracers, but is 3D with walls effects for small particles. The disc is half-filled and its axis horizontal. Rotation is obtained with a motor at 2 speeds: 0.033~s$^{-1}$ corresponding to successive 
avalanches and one speed between 0.8 and 0.16~s$^{-1}$ chosen to obtain a 
continuous flow and a plane free surface. We observe through the walls what 
species is present at the surface of the flowing layer. One limitation is 
the formation of plugs when tracer size is close to the drum thickness \cite{moi}. Strong electrostatic 
effects happen, especially with small Diakon and Wood Nut particles. 
Consequently, results are not accurate and we focus on 
the general behavior. The criterion ``being visible on the surface during the flow" gives 2 identical sets of results for the 2 different speeds: the segregation process in a continuous flow or in a single 
avalanche is the same. We thus only draw 1 point which represents the type of segregation for each couple 
($\rho_2/\rho_1$, $d_2/d_1$) [Fig.~\ref{RCdisc}, discussed in Sec.~\ref{RC}].

\subsubsection{Formation of a pile}

\begin{figure}[htbp]
\center
\includegraphics[width=7.8cm]{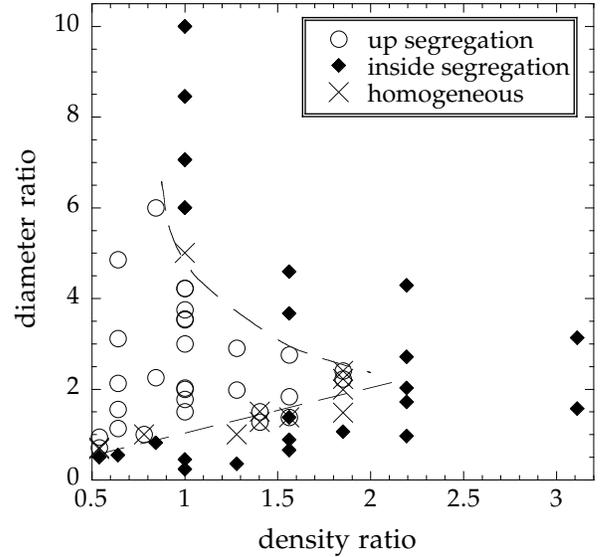}
\caption{Surface segregation in {\bf piles}: tracers ({\Large{$\circ$}}) making a bottom ring on the pile, ($\blacklozenge$) embedded at the top of the pile, ({\large $\times$}) piles looking homogeneous. See caption Fig.~\ref{RCdisc}.}  
\label{RCtas}
\end{figure}

The heap formed beneath a funnel exhibits features of segregation \cite{jullien90, boutreux98, moi}. Here, they fully result from processes taking place in the flow along the pile slope: dynamical sieving, rolling of large beads on a relatively smooth surface \cite{riguidel94, celine} and push-away process \cite{moi}. No segregation happens in the funnel \cite{arteaga} because the mixture slips along the walls, contrary to \cite{silo}.
 The particles at the surface of the flowing phase go further than the others and accumulate in a ring at the pile base. Consequently, 
an up-segregation of tracers in the flow produces a basal ring of tracers, contrary to an intermediate level segregation for which tracers stop on the pile sides, and to a bottom segregation for which tracers stop near the summit (pictures in~\cite{moi}).
The funnel contains 250 to 850~g of a homogeneous mixture and is placed at 12~cm above a rough plane. Fluxes range between 10 and 30~g/s depending on 
particles used. One limitation is the occurrence of blocking of the funnel for $d_2$$>$5~mm. The flow is continuous along the pile slope in the first stage, then proceeds as axisymmetric avalanches when the pile 
gets bigger. The flowing layer is thin, and vanishes to zero along the slope when the pile 
gets bigger, allowing the possibility to see buried particles. 
 Because buried tracers are visible, the chosen criterion is the formation of a basal ring of tracers. It gives the up (ring) and inside (no ring) segregation fields [Fig.~\ref{RCtas}, discussed in Sec.~\ref{RC}]. 

\subsubsection{Flow along a slope}\label{chenal}

\begin{figure}[htpb]
\center
\includegraphics[width=7.8cm]{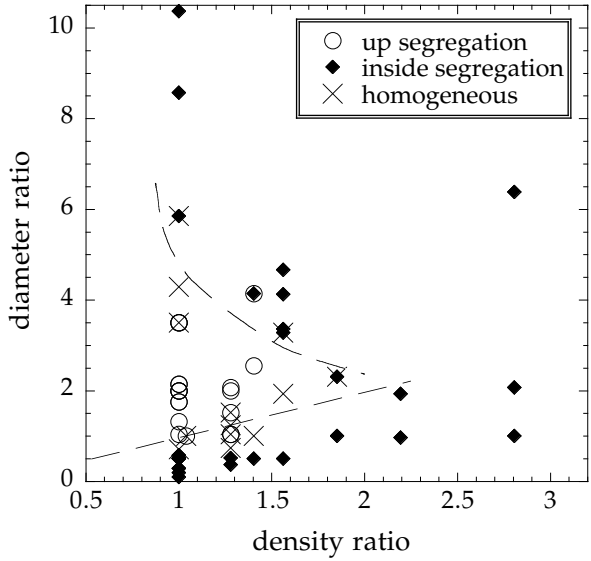}
\caption{Surface segregation in the {\bf channel}: tracers ({\Large{$\circ$}}) cover the surface, ($\blacklozenge$) are embedded in the deposit or at the bottom, ({\large $\times$}) are both inside and at the surface (homogeneous). See caption Fig.~\ref{RCdisc}.}
\label{RCchenal}
\end{figure}

Chute flows have been already studied for mixtures of different size or/and density particles \cite{savagelun, khakhar99chute, Dolgunin95, Dolgunin, moi}. We choose to study frictional flows, and avoid collisional flows.
The chute is a 1~m long, 6~cm wide channel inclined with an angle of about 
27$^\circ$ for beads (26.5-30$^\circ$), and up to 33.5$^\circ$ for angular particles, chosen to produce frictional flows [Fig.~\ref{devices}]. The bottom is made rough by gluing 
Glass beads or angular fragments on it, whose size is those of the bed particles. The starting box is filled with $\sim$1~kg of a 
homogeneous mixture,  then opened, producing a thick flow with a flux of about 400~g/s. The flow
stops where the channel is horizontal. Observations of the flow and 
of the internal structure of the deposit show that the deposit results of 
a strong deformation of the flow. However, the surface of the flow 
stays on the surface of the deposit: there is little doubt in 
diagnosing the location of a species above or under the surface of the flow by looking only at the deposit. We took as criterion whether the tracers are visible or not at the surface deposit [Fig.~\ref{RCchenal}, discussed in Sec.~\ref{RC}]. 

\begin{figure}[htpb]
\center
\includegraphics[width=7.8cm]{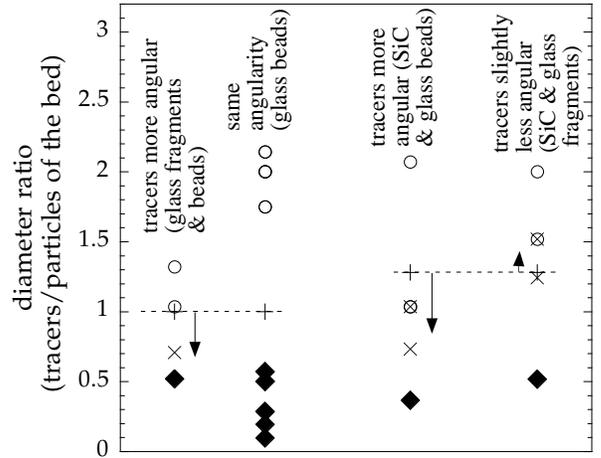}
\caption{Channel experiments for various angularity contrasts: ({\Large{$\circ$}}) tracers at the surface, ($\blacklozenge$) tracers buried in the deposit, ({\large $\times$}) homogeneity, ({\large +} and dashed line) expected homogeneity for particles of the same shape \cite{alonso}. See caption Fig.~\ref{RCdisc}. When angular and spherical particles of the same `size' and density are mixed, there is an up-segregation of angular particles. The use of more angular particles as tracers (resp. as bed particles) seems to shift down (resp. up) the size ratio for which the mixture remains homogeneous (indicated by arrows). SiC are tracers: $\rho_2/\rho_1$$=$1.28 (right); 1 (left).}
\label{forme}
\end{figure}

The particle shape influences the segregation pattern and the ratio values corresponding to the limit [Fig.~\ref{RCchenal}]. The relative sizes in the 3 axes of particles allow to estimate their angularity: typically a factor 1-2 between smaller and larger sizes for Sand, a factor 2-3 for SiC, Sugar and Wood fragments, a factor 5 for Glass fragments. When mixed, all these particles give several angularity contrasts [Fig.~\ref{forme}]. First, we found that for ``equal sized" particles (same sieving interval) angular particles segregate at the surface (Glass fragments and beads ({\Large{$\circ$}}) for size ratio 1). These results seem in accordance with shaking simulations of elongated particles \cite{Abreu} which show different segregations depending on the elongation. 
Second, when tracers are more angular than bed particles (SiC or Glass fragments in beads), the homogeneous state seems to be obtained for smaller size ratios than expected \cite{alonso}. For Glass, the shift in size ratio extends roughly from 1 to 0.7: Glass fragments behave as beads approximately 1.4 bigger. When tracers are less angular (SiC tracers in Glass fragments), the homogeneous state seems to appear for greater tracers than expected. These deviations suggest that the more angular particles are pushed toward the surface, although the gap between the experiments is too large to conclude, except for size and density ratios 1. It is in accordance with the 4 ({\large $\times$}) under (resp. 3 ({\large $\times$}) above) the bottom limit [Fig.~\ref{RCchenal}] corresponding to tracers more (resp. less) angular than bed particles. 
But, for particles whose 3 dimensions are close, the effect of shape is small explaining why data regroup in two distinct zones [Figs.~\ref{RCdisc}-\ref{RCchenal}]. 

\subsection{Recursive limit between the fields of segregation}\label{RC}

On each set of results, two distinct domains can be defined in ($\rho_2/\rho_1$; $d_2/d_1$) coordinates [Figs.~\ref{RCdisc}-\ref{RCchenal}] showing that the segregation pattern is a function of size and density ratios rather than diameter and density. 
The results of the 3 systems are very similar: the same dashed limit is plotted on the 3 graphs. The segregation process does not depend on the geometry inducing the frictional flow.
The top part of the limit has been chosen as a compromise using all sets of data. The bottom part is the line $\rho_2/\rho_1$=$d_2/d_1$ obtained by \cite{alonso} for homogeneous mixtures at low tracer fractions.
We notice that processes happening when shaking or shearing a granular medium are different: our 3 identical sets of results are different from those of a vertically shaken box~\cite{breu}.

On each graph, the limit passes between the two segregation types, or where there is no observed segregation (marked as homogeneous). The limit shape is recursive, {\it i.e.} it is not a curve of a single-valued function from [$\rho_2/\rho_1$] to [$d_2/d_1$]; the area of visible tracers ({\Large{$\circ$}}) is surrounded by those of buried tracers ($\blacklozenge$). 
Consequently, a line $\rho_2/\rho_1$=constant can cross twice the limit, depending on the $\rho_2/\rho_1$ value. For these ratios, small and very large tracers are under the surface, and medium size tracers are on the surface. Consequently, the mean position of a growing tracer evolves up then down, in an analogous way of the $r/R$ curves [Figs.~\ref{r/Rpropyl}-\ref{r/Rplomb}]. This implies that the same level can be reached by two tracers of the same density, with different diameters. We deduce that the shape of the $r/R$ vs $d_2/d_1$ curves [Figs.~\ref{r/Rpropyl}-\ref{r/Rplomb}] is not due to the use of excessively large beads constituting the bed in the 3D drum (Sec.~\ref{cloche}). For $\rho_2/\rho_1$$\gtrsim$1.8 (1.6-2.2), tracers are always hidden. In the 3D drum, these density ratios correspond to tracers which never go higher than $r/R$=0.7 (between 0.84-0.6) whatever their sizes are. It implies that positions visible in surface correspond to approximately $r/R$$>$0.7 (strictly $r/R$$>$0.84) and that the level of the limit might be $r/R$=0.7 in the drum.

\begin{figure}[htpb]
\center
\includegraphics[width=9.37cm]{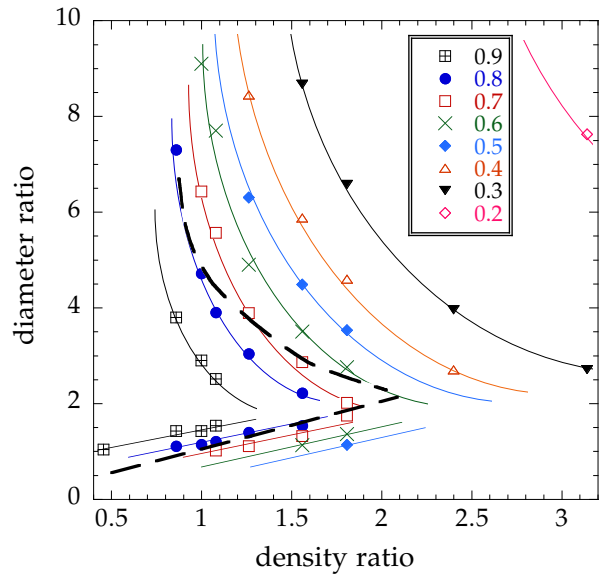}
\caption{Curves linking mixtures in which tracers segregate at the same $r/R$ in the 3D drum ($r/R$ values in the inset). See caption Fig.~\ref{RCdisc}. Symbols are values taken from the fits \cite{fit} of $r/R$ vs $d_2$/$d_1$, lines are just guides for eyes. Iso-$r/R$ 0.7 and 0.8 are close to the limit (thick dashed curve) between segregation fields of pile/disc/channel experiments.}
\label{iso}
\end{figure}

Vice-versa, the $r/R$ vs $d_2$/$d_1$ curves [Figs.~\ref{r/Rpropyl}-\ref{r/Rplomb}] suggests that not only the level of the limit but all levels in the flow have a recursive shape when represented in ($\rho_2/\rho_1$; $d_2/d_1$) coordinates. The intersections of the $r/R$ vs ($\rho_2/\rho_1$, $d_2/d_1$) surface (Sec.~\ref{drum}) by planes $r/R$=constant have effectively recursive shapes [Fig.~\ref{iso}]. Moreover, the limit defined from disc-pile-channel data, is close to curves $r/R$=0.7 and 0.8: it corresponds roughly to one iso-location, and it is close to $r/R$=0.71 which represents the homogeneity (Sec.~\ref{cloche}).

Moreover, the 3D drum experiments show that the iso-position curves are not symmetrical in terms of segregation intensity: each lower part, close to line $\rho_2/\rho_1$$=$$d_2/d_1$, corresponds to large standard deviations, and each upper part to small standard deviations. This fact is compatible with cross-sections of the channel deposit, which look rather homogeneous on the lower part of the limit, and rather segregated at an intermediate level close to the surface for the upper part. These observations are only visual, but seem to confirm that large size ratios enhance the focus of the segregation pattern, whereas both size and density ratio determine the mean position of tracers.

\section{Conclusions}

We found that large tracers segregate either at the bottom, at the surface or, in most cases, at an intermediate level in the flow. This creates well-defined ring patterns of tracers in a half-filled rotating drum. Reversing between surface and bottom segregation is also observed in chute flow and pile building where the granular matter is subject to a frictional flow. 

In the drum device, the radius of the ring varies continuously with both the density and the size ratios between tracers and beads of the bed. A difference of size has an equivalent effect on the tracer segregated position than a difference of density. Compared to the size segregation of beads of the same density, the progressive sinking of large tracers when increasing the size ratio is enhanced (resp. delayed) for denser (resp. lighter) tracers. 

These facts are interpreted by the existence of two segregation processes affecting the tracers: a geometrical effect (dynamical sieving process) and a mass effect (push-away process). The push-away effect makes heavy beads go towards the bottom of the flow. The geometrical effect drives large beads towards the free surface. We could have expected that the density difference between the beads is responsible for one effect, and that the size difference is responsible for the other one. But, a size difference introduces the two effects of segregation (geometrical and mass), while a density difference implies only the mass effect. When tracers segregate on a ring, the two effects exactly balance at a well-defined vertical level inside the flow. This level is the equilibrium location for these tracers. The segregation of the tracers appears consequently to the regrouping of all the identical tracers at the same equilibrium location. 

In the size segregation between beads of the same density, the geometrical effect is dominant at small size ratios (large beads segregate at the surface, as usually observed); the mass effect is dominant at very large size ratios (large beads segregate at the bottom); geometrical and mass effects balance for intermediate ratios (large beads segregate at an intermediate level). The variation of the relative intensities of both effects explains the continuous progressive reversing of the position of large tracers when increasing the size ratio.

Because the balance happens at one particular level when a ring forms, at least one of the two segregation effects depends on the vertical coordinate in the flow.
The evolution of the ring pattern gives two types of information: (1) the focus of the pattern determines the intensity of the segregation process. It is observed to be strong for high size ratios, weak for small size ratios, whatever the density ratio is. (2) The variation of the radius of the ring shows that different mass effects can be in balance with geometrical effects due to the same size ratio, but acting at different levels in the flow. 
Data on focus and radius allow to conclude that the geometrical effect depends strongly on the size ratio, as expected, but also on the vertical location in the flow. While the mass effect depends mostly on the size and density ratios.

Consequently, the two effects (mass and geometrical) implied in the {\it size segregation of beads of equal density} can be separated when using the whole set of data acquired on the locations of {\it segregated beads of different densities and sizes}:
(1) Considering two tracers of different sizes and densities at the same location, the difference of geometrical effect due to the size difference is exactly balanced by the difference of mass effect. (2) Considering two tracers of the same size, at two different locations, which is possible if they have different densities, the difference of geometrical force due to the variation of the location is exactly balanced by the difference of mass effect. By modeling the mass effect, the geometrical effect can be calculated in function of both location and size ratio. This is currently under development in a model of the geometrical effect responsible for the segregation \cite{futur}. 
\\

\textbf {Acknowledgements :} We thank E.~Cl\'ement for advises, drum and some beads, B.~Dalloz-Dubrujeaud for Zirblast beads, U.~D'Ortona for many fruitful discussions.


\begin{thebibliography}{99}
\bibitem{fit}
The function (no physical meaning) used to establish the lines is:  $r/R=m_1+m_2/(exp((d_2/d_1-m_4)*m_3)+exp((-d_2/d_1+m_4)*m_5))$, 
where $m_1$, $m_2$, $m_3$, $m_4$, $m_5$ are adjustable parameters. Data of Fig.~\ref{iso} are obtained from these fits. Two successive functions have been used when $r/R$ saturates at the periphery.
\bibitem{lead}
Lead tracers $r/R$ are sometimes slightly above Steel tracers $r/R$. This may result of a little sinking of Lead tracers in the solid phase of the bed due to very large $\rho_2/\rho_1$ as previously found by \cite{ristow94}.
\bibitem{fit2} 
The function (no physical meaning) used to establish the lines is:   $r/R=n_1-n_2*tanh((\rho_2/\rho_1-n_4)*n_3)$ where $n_1$, $n_2$, $n_3$, $n_4$ are adjustable parameters.
\end{thebibliography}
\end{document}